\documentclass[12pt]{article}

\usepackage{xspace}

\newcommand{\ie}{\textit{i.e.}\xspace}

\begin{document}

\newcommand{\nc}{\newcommand}

\nc{\beq}{\begin{equation}}
\nc{\eeq}{\end{equation}}
\nc{\beqa}{\begin{eqnarray}}
\nc{\eeqa}{\end{eqnarray}}
\nc{\lsim}{\begin{array}{c}\,\sim\vspace{-21pt}\\< \end{array}}
\nc{\gsim}{\begin{array}{c}\sim\vspace{-21pt}\\> \end{array}}
\nc{\el}{{\cal L}} \nc{\D}{{\cal D}}

\nc{\R}[1]{\ensuremath{R_{#1}}}
\nc{\dimlessR}[1]{\ensuremath{\hat{R}_{#1}}}
\nc{\gs}{\ensuremath{g_{s}}}
\nc{\gYM}{\ensuremath{g_{YM}}}
\nc{\ls}{\ensuremath{l_{s}}}
\nc{\lp}{\ensuremath{l_{P}}}
\nc{\w}{\ensuremath{\omega}}
\nc{\z}{z}

\renewcommand{\thepage}{\arabic{page}}

\begin{titlepage}
\begin{center}
\bigskip

 \hfill YCTP-P3-01\\

 \hfill hep-ph/0105021\\

\bigskip
\bigskip
\vskip .5in

 {\Large \bf Casimir Energy and Radius Stabilization

\vskip .1in

in Five and Six Dimensional Orbifolds}

 \vskip .8 in

 {\large Eduardo Pont\'{o}n} and
  {\large Erich Poppitz}\footnote{Address after
 July 1, 2001: Department of Physics,
 University of Toronto, 60 St George St.,
 Toronto, ON M5S 1A7, Canada}

 \vskip 0.2in

 \vskip 0.2in {\em
  Department of Physics

  Yale University

 New Haven,
   CT 06520-8120, USA}

\end{center}

\vskip .4in

\begin{abstract}

\end{abstract}

We compute the one-loop Casimir energy of gravity and matter
fields, obeying various boundary conditions, in 5-dimensional
$S^1/Z_2$ and 6-dimensional $T^2/Z_k$ orbifolds. We discuss the
role of the Casimir energy in possible radius stabilization
mechanisms and show that the presence of massive as well as
massless fields can lead to minima with zero cosmological constant.
In the 5-d orbifold, we also consider the case where kinetic terms
localized at the fixed points are not small. We take into account
their contribution to the Casimir energy and show that localized
kinetic terms can also provide a mechanism for radius
stabilization. We apply our results to a recently proposed
5-dimensional supersymmetric model of electroweak symmetry breaking
and show that the Casimir energy with the minimal matter content is
repulsive. Stabilizing the radius with zero cosmological constant
requires, in this context, adding positive bulk cosmological
constant and negative brane-tension counterterms.

\end{titlepage}

\setcounter{page}{1}

\renewcommand{\thefootnote}{\#\arabic{footnote}}
\setcounter{footnote}{0}

\baselineskip18pt

\section{Introduction and summary.}

The idea that there are extra spatial dimensions in which gravity
and, possibly, some or all matter fields can propagate has been the
subject of renewed interest in the last few years. The Kaluza-Klein
and  ``braneworld"  ideas combined have been used to reformulate
and address every conceivable problem of elementary particle
physics \cite{Kaluza} -- \cite{Randall:1999vf}.

One of the main theoretical issues in theories with extra
dimensions is that of determining their size. In the absence of a
stabilization mechanism, the Casimir energy tends to either inflate
or contract the extra dimensions, as has been known since the work
of refs.~\cite{Appelquist:1983vs, Appelquist:1983zs}. The Casimir
energy of various braneworld compactifications has also received
attention recently \cite{Fabinger:2000jd, Garriga:2000jb,
Toms:2000vm, Goldberger:2000dv, Hofmann:2000cj, Brevik:2000vt}.

In this paper, we focus on the Casimir energy and its role in
radius stabilization in a particular class of Kaluza-Klein
compactifications not discussed   until recently---field-theory
orbifold compactifications.\footnote{We note that all field theory
orbifolds  discussed in this paper have some periodic fermions,
hence the nonperturbative instability of the Kaluza-Klein vacuum
towards decay into ``nothing,'' pointed out in \cite{Witten:1982gj}
and recently discussed in \cite{Fabinger:2000jd} is absent.}
Orbifolds are very useful tools for projecting out unwanted
massless modes and/or breaking (super)symmetries. Several
interesting phenomenological models using field theory orbifolds
were proposed recently \cite{Barbieri:2001vh, Arkani-Hamed:2000hv}.

In  five dimensions, an orbifold   was used to construct a
supersymmetric model of (calculable!) electroweak symmetry breaking
\cite{Barbieri:2001vh}, predicting a light Higgs (related models
were considered before in
Refs.\cite{Delgado:1999qr,Antoniadis:1999sd}). The realization of
supersymmetry in this model is quite different from the usual
minimal supersymmetric standard model---the theory below the
compactification scale is not supersymmetric, while superpartners
(the lightest being the stop) as well as mirror particles appear
near that scale. In the six-dimensional case, a nonsupersymmetric
$T^2/Z_2$ orbifold construction was employed to build a
higher-dimensional  composite-Higgs model of electroweak breaking
\cite{Arkani-Hamed:2000hv}. The issue of radius stabilization was
not addressed in refs.~\cite{Barbieri:2001vh,
Arkani-Hamed:2000hv}---the Higgs-dependent part of the Casimir
energy played a crucial role in the analysis of electroweak
symmetry breaking in \cite{Barbieri:2001vh}, but the radius was
assumed to be fixed.

Originally,  orbifolds were invented in string theory
\cite{Dixon:1985jw, Dixon:1986jc} as a tool for constructing new
consistent string backgrounds. Modular invariance of the worldsheet
orbifold CFT and/or tadpole cancellation
 \cite{Bianchi:1990yu, Bianchi:1991tb, Gimon:1996rq}
place severe constraints on their consistency as fundamental
theories. The field theory orbifolds that we discuss in this paper
cannot, at least at present,  be derived from known string
constructions. Nevertheless, their success with electroweak
symmetry breaking is appealing enough to warrant further study. We
will only demand that the orbifold theories be consistent as
low-energy effective theories (hence the requirement that the
theory of the zero modes be anomaly free; for a recent discussion,
see \cite{Arkani-Hamed:2001is}), valid up to some energy cutoff
scale. Consistency of the field theory description demands then
that the cutoff be  (at least) an order of magnitude or so larger
than the inverse compactification length scale.

When performing loop calculations in the orbifold field theory, the
orbifold boundary conditions lead to extra divergences  at the
orbifold fixed points, see, e.g., \cite{Birrell:1982ix}. To cancel
these, new terms localized at the fixed points have to be
introduced in the lagrangian. These localized terms can be kinetic
and mass terms, as well as interaction terms \cite{Georgi:2001ks,
Goldberger:2001tn}. Their coefficients are additional parameters of
the orbifold theory. It is known \cite{Dvali:2000hr}, that these
``brane-localized'' terms can affect the spectrum of the theory and
consequently  the Casimir energy. It is then natural to expect that
they can  play a role in the mechanism of radius stabilization.

We will first work in an approximation  neglecting the
contributions of brane-localized kinetic and mass terms to the
Casimir energy. This is a valid approximation if the coefficients
of brane-localized kinetic and mass terms are of the order of the
loop-generated values (consistency and naturalness demand that the
tree-level coefficients be at least as large as the loop-induced
values \cite{Chacko:2000hg}). We will then generalize the
calculations to the case of larger brane-localized kinetic and mass
terms.

In Section \ref{5dsetup},  we describe the five dimensional
$S^1/Z_2$ setup. We then  present, in Section
\ref{generaldiscussion}, a general discussion of the divergences in
the one-loop Casimir energy, the counterterms required for their
cancellation, and the contributions to the radius potential from
the Casimir energy and counterterms. The  discussion of Section
\ref{generaldiscussion} applies equally well to the 6-d case.

Motivated by the 5-d and 6-d models, mentioned above, we calculate
the Casimir energy in  $S^1/Z_2$, in Section
\ref{5dcasimirmassless}, and $T^2/Z_k$ compactifications, in
Section \ref{6dcasimir}. We compute the gravitational contribution,
as well as those of even or odd massless matter fields obeying
periodic or antiperiodic boundary conditions. In Section
\ref{5dcasimirmassive}, we also calculate the Casimir energy of
  fields with bulk mass terms in the 5-dimensional case and
  discuss a mechanism for stabilizing the radius with
  massive fields. The computations
of Sections \ref{5dcasimirmassless}, \ref{5dcasimirmassive}, and
\ref{6dcasimir}
   are sufficiently general to be useful when
considering radius stabilization in concrete five and six
dimensional models.

In Section \ref{branecasimir}, we include the contribution of the
brane kinetic terms to the Casimir energy in the $S^1/Z_2$
orbifold. We show that the brane kinetic terms,  if they are
sufficiently large, can lead to radius stabilization at a size
bigger than the cutoff length scale. For example, a value
consistent with ``naive dimensional analysis" (NDA) can yield a
radius several times larger than the cutoff
 length. NDA arguments \cite{Chacko:2000hg}
constrain the coefficients of brane-localized kinetic terms (with
dimension of length) to be several times the cutoff length scale.

A much larger value of the brane-localized kinetic terms
  can yield a radius much bigger than the cutoff length
scale. To investigate the viability of such a scenario as an
effective low-energy theory, in Section \ref{branepropagator} we
ask whether the theory has a consistent perturbative expansion if
the brane kinetic terms are significantly larger than the cutoff
length scale. We investigate in detail the properties of the
Green's function  with the brane kinetic terms included and find
that the brane-localized divergences are weaker, while bulk
divergences are as in the theory without the brane kinetic terms;
 we note that this is consistent with the results of
refs.~\cite{Dvali:2000hr, Dvali:2001gm, Carena:2001xy}. The
analysis of this Section strengthens the case for having values for
the brane kinetic terms larger than those implied by
NDA.\footnote{This does not imply that the NDA arguments are
incorrect. The main assumption of NDA---that there is a single
fundamental length scale---does not hold when the brane kinetic
terms take values larger than the cutoff length; we are merely
saying that such a scenario  can yield a consistent low-energy
theory.}

Finally,   in Section \ref{applications}, we apply our results to
the 5-d supersymmetric models of electroweak breaking
\cite{Barbieri:2001vh} (see also \cite{Arkani-Hamed:2001mi}). We
show that the Casimir energy contribution of the massless fields to
the radius potential is repulsive (here we treat  brane kinetic
terms as small). We show that by fine-tuning the 5-d cosmological
constant and ``brane tension'' counterterms (as discussed in
Section \ref{generaldiscussion}), an acceptable  minimum for the
radius  with vanishing cosmological constant can be achieved. The
sign of the bulk cosmological term required to have such a minimum
is positive, while the brane tensions need to be negative. It  thus
appears that the required bulk counterterm is not supersymmetric,
at least in the simplest 5-d supergravities. This might present a
problem, because a non-supersymmetric counterterm would introduce
an additional source of supersymmetry breaking and potentially
affect the predictions of the model. Thus, deciding whether this,
or any other, stabilization mechanism is viable requires fully
embedding the model in 5-d supergravity; this is a   problem  that
we leave for future work.

We present details of the calculation of one of the Casimir sums in
an Appendix (other sums in the paper, where indicated, are computed
similarly). We end with concluding remarks and a discussion of some
open issues in Section \ref{conclusions}.

\section{Casimir energy on orbifolds.}
\label{5dcasimir}

In this Section, we discuss the general issue of the Casimir energy
on orbifolds, in the framework of the five dimensional $S^1/Z_2$
example. We begin by  introducing the setup and notation in Section
\ref{5dsetup}. In Section \ref{generaldiscussion}, we discuss the
limits of applicability of the Casimir energy calculation,  the
divergences, and the counterterms required for their cancellation.
The coefficients of these counterterms can not be computed from the
low-energy effective theory alone---they are to be treated as
parameters of the theory and are fixed by imposing normalization
conditions on the potential for the radius. Sections
\ref{5dcasimirmassless}, \ref{5dcasimirmassive} contain the results
of the calculation of the Casimir energy contribution of the
gravitational field as well as of various massless and massive
fields. In Section \ref{5dcasimirmassive} we also discuss a
possible stabilization mechanism using the Casimir energy of
massive fields.

\subsection{The  five dimensional $S^1/Z_2$ example.}
\label{5dsetup}

The general parameterization of the interval of an $S^1$
compactification is as follows, see, e.g.,
\cite{Appelquist:1983vs}:
\beq
\label{5dinterval} d s^2 = \phi^{-{1\over 3}}
\left( g_{\mu\nu}+ A_\mu A_\nu \phi \right)~ d x^\mu d x^\nu  +
2 \phi^{2\over 3} A_\mu  d x^\mu d y  + \phi^{2 \over 3} d y^2~.
\eeq
Here the four-dimensional indices are denoted by $\mu$ and the
five-dimensional coordinate $y$ (taken to have dimension of length)
is assumed to change between $-L$ and $L$ (until the physical
radius is fixed, the scale $L$ is a completely arbitrary length
scale). For a general fluctuating background the fields $g_{\mu
\nu}$, $A_{\mu}$, and $\phi$ can depend on $x^\nu$ and $y$. The
$Z_2$ orbifold is obtained after identifying points on the circle
related by a
 reflection in the fifth coordinate, $y
\simeq -y$.
The invariance of the interval (\ref{5dinterval}) under the $Z_2$
symmetry determines the transformation properties of the fields:
$g_{\mu\nu}(y) = g_{\mu\nu} (-y)$, $A_\mu (y) =
- A_\mu (-y)$, and $\phi(y) =
\phi(-y)$.
Since the field $A_\mu$ is odd under the $Z_2$ symmetry, it can not
have a zero mode.

The parameterization of the metric in (\ref{5dinterval}) is
convenient,  because in the four-dimensional effective field theory
of the zero modes it gives rise to four-dimensional Einstein
gravity  with metric tensor $g_{\mu\nu}(x)$, coupled to a
dilaton\footnote{We will occasionally call this field a `radion.'}
$\phi(x)$ and, before the orbifold projection, an abelian gauge
field $A_\mu(x)$. Since the $Z_2$ orbifold forbids the appearance
of a zero mode of $A_\mu$, we omit this field in what follows. The
zero-mode effective theory is valid below the scale of the mass of
the lowest Kaluza-Klein excitation, i.e. at energies below
$(\phi^{1/3} L)^{-1}$; note that the physical size of the extra
dimension is $\phi^{1/3} L$. More precisely, the five dimensional
Einstein action, evaluated on a background (\ref{5dinterval}) with
the fields $g_{\mu\nu}(x),
\phi(x)$, dependent only on $x^\mu$ is given by:
\beq
\label{5dreduction}
M_5^3 \int d^5 x \sqrt{G} R(G) = M_4^2 \int d^4 x \sqrt{g} \left[~
R(g) + {1\over 6} {\partial_\mu \phi ~ \partial_\nu \phi
~g^{\mu\nu}
\over \phi^2}~\right]~.
\eeq
In the above formula $G$ denotes the five dimensional metric tensor
that can be read off eqn.~(\ref{5dinterval}), with $A_\mu = 0$ and
the rest of the fields only dependent on $x^\mu$, while the four
and five dimensional Planck scales are related by $M_4^2 = L
M_5^3$; in the orbifold theory we only integrate over the
fundamental region $0
\le y \le L$.\footnote{To avoid confusion, we note that
 the relation between four dimensional and five dimensional
Planck scales involves the physical size of the orbifold,
$\phi^{1/3} L$,  instead of the arbitrary scale $L$; however, the
form (\ref{5dreduction}) is more convenient before fixing $\langle
\phi \rangle$.}

\subsection{Casimir energy, divergences, and counterterms.}
\label{generaldiscussion}

Our goal is to study the generation of a potential $V(\phi)$---and
the possible existence of a minimum---for the dilaton field $\phi$
due to quantum effects of the nonzero modes of the fields $g, A,
\phi$, as well as  to  quantum contributions to the Casimir energy
of   matter fields (massless or not) that might be present.

We will perform the calculation of the Casimir energy around a
constant background $g_{\mu\nu} =
\eta_{\mu\nu}$, $A_\mu = 0$, $\phi = const.$, with $\eta_{\mu\nu}$--the
Minkowski metric. Dynamical issues, such as the time evolution of
the background as a backreaction to the Casimir energy, can then
only be studied for time intervals such that the deviation of the
metric from the assumed constant background is small. In this
paper, we are interested in the existence of static stable minima
with vanishing four-dimensional cosmological constant. Other
interesting issues, such as the cosmological evolution of the
background, are left for future work.

We do not attempt to say anything about the cosmological constant
problem here. To achieve vanishing cosmological constant at the
minimum, we will resort to the fact that the  computation of the
Casimir energy is plagued with divergences, whose cancellation
requires adding counterterms to the action (\ref{5dreduction}).
Divergences are short-distance phenomena and the  counterterms
needed to cancel them are local terms, respecting the
short-distance symmetries of the theory. The Casimir energy, on the
other hand, is a global effect, depending on the topology and
boundary conditions of the compactification, and can not be
described by a local term preserving the short-distance symmetries.
For example, in the $S^1$ compactification, (bulk) counterterms
(using a generally covariant regulator) should respect 5-d general
covariance and should not depend on the fact that at large
distances it is broken by the compactification. For the constant
flat metric background of interest to us, there is only one
divergent counterterm---the 5-d cosmological constant term (this
is, strictly speaking, true only in the unorbifolded case, see
discussion below). This term is:
\beq
\label{cccounterterm}
~\alpha~ \int d^5 x \sqrt{G}   =  \alpha  L~ \int d^4 x~
\phi^{- 1/3}~,
\eeq
and thus contributes a potential  $\sim \phi^{-1/3}$ to the
four-dimensional effective action (\ref{5dreduction}).  We will
treat its  coefficient $\alpha$ (of dimension (mass)$^5$) as a
parameter to be fixed by the normalization conditions of the
potential:
\beq
\label{normalizationconditions}
V(\phi)
= V^\prime (\phi) = 0~.
\eeq
 We note that for  values of $\phi$, such that eqns.~(\ref{normalizationconditions}) hold,
  the metric background is flat and thus the calculation of the Casimir
energy---which assumed that---is self-consistent.

It is clear that if only massless fields are present and the only
scale in the problem is $M_5$, the potential for $\phi$ due to the
 massless fields' Casimir energy  is monotonic in $\phi$ and minimized either at
$\phi
= 0$ or $\infty$, depending on the matter content of the theory.
We will see that the massless fields' Casimir contribution to the
effective action (\ref{5dreduction}) is proportional to:
\beq
\label{masslesscasimirpotential}
{1\over L^4}~\int d^4 x ~\phi^{-2}.
\eeq
 If the potential (\ref{masslesscasimirpotential}) is minimized for
 $\phi \rightarrow 0$, one has to resort to strong curvatures
 or other effects beyond the
reach of the effective theory  to stabilize the radius and   the
effective theory description of the low-energy physics is not
valid. If, on the other hand, the minimum is at $\phi
\rightarrow
\infty$ one has a
typical runaway potential (exponential, if the kinetic terms are
rendered canonical by a field redefinition). This potential may
provide for an acceptable  form of  quintessence.  Our calculations
show that in the five-dimensional models with the supersymmetric
standard model in the bulk (with the minimal matter content) it is
the second possibility that is realized.

Adding the cosmological constant counterterm $\sim
\phi^{-1/3}$, eqn.~(\ref{cccounterterm}), changes the shape of the potential
(\ref{masslesscasimirpotential})  and can generate a minimum.
However, it is impossible to have a vanishing cosmological constant
at the minimum with only the  two contributions
(\ref{masslesscasimirpotential}) and (\ref{cccounterterm})  to the
potential. The nonzero cosmological constant at the minimum renders
the calculation inconsistent.

Clearly, the resolution is to introduce another scale in the
problem. In what follows we consider the possible ways to do this.

The first, ``nonminimal," way is to introduce additional fields in
the theory and is applicable equally well to the unorbifolded
theory. One can, for example,
   include a field with mass smaller than the cutoff of the
   five dimensional theory. In Section \ref{5dcasimirmassive},
we calculate the Casimir energy of massive fields and show that
upon adding the three contributions---massless, massive, and
cosmological constant counterterm---it is possible to achieve a
stable minimum for $\phi$ with vanishing cosmological constant,
 thus rendering the  calculation self-consistent. The physical
 size of the extra dimension then is typically of order the inverse
 mass of the field. However, we will see that this mechanism,
 though in principle viable, can not be implemented in
 realistic models with the (supersymmetric) standard model fields in
 the five-dimensional bulk.

A second possibility---we call it ``minimal," since it does not
involve introducing new fields solely for the purpose of radius
stabilization---is to explore the fact that we are compactifying
not on a smooth circle, but on an orbifold. The Green functions
$G(x-x^\prime; y, y^\prime)$ of fields with orbifold boundary
conditions have, in addition to the usual short-distance
singularity when $x^\mu
\rightarrow x^{\mu\prime}, y \rightarrow y^\prime$, a singularity
localized at the fixed points of the orbifold $y= y^\prime = 0, L$
(see for example ref.~\cite{Birrell:1982ix}, where the Green
function of a scalar field in four dimensions with Dirichlet
boundary conditions at a fixed two-plane is discussed). These
singularities lead to  divergent terms localized at the fixed
points; for recent discussions see refs.~\cite{Georgi:2001ks,
Goldberger:2001tn}.

For example, since the Casimir energy is the expectation value of
the energy momentum tensor, related to the Green function (e.g., of
a scalar field) by:
\beq
\label{tmunu}
\langle T_{MN}(X) \rangle
= (\partial_M\partial^\prime_N + \ldots) G(X, X^\prime)\big\vert_{X
\rightarrow X^\prime}~,
\eeq
(here $X$ denotes both $x^\mu$ and $y$), the singularities of
$G(X,X^\prime)$ at the orbifold fixed points lead to divergent
contributions to the Casimir energy localized at the fixed points,
in addition to the 5-d bulk cosmological constant term mentioned
above. Canceling these extra divergences  requires adding
counterterms localized  at the fixed points (i.e. ``brane
tensions"). We will treat the coefficients of these terms   as
parameters, which can not be determined in the low-energy theory.
(We have to assume that the ``brane tensions" are small enough in
order to ignore the warping they typically produce; we will comment
on this in our discussion in Section \ref{applications}.) For the
metric background (\ref{5dinterval}), these localized terms on the
$S^1/Z_2$ orbifold are of the form:
\beq
\label{branetensions}
\beta \int d^5 x \delta(y) ~\sqrt{\tilde{g}} = \beta \int d^4 x ~
\phi^{-2/3}~,
\eeq
where $\tilde{g}$ denotes the induced metric at $y=0$, and a
similar term at the $y = L$ fixed point. Thus the localized terms
scale differently from the terms due to the cosmological constant
counterterm (\ref{cccounterterm}) and the massless fields' ``bulk"
Casimir energy (\ref{masslesscasimirpotential}). We will show that
in some cases it is possible to achieve a (local) minimum of
$V(\phi)$ with vanishing cosmological constant. The physical size
of the extra dimension is then set by the coefficients of the
tension terms.

The singularity of the Green function leads also to other
divergences, for example to kinetic terms localized at the orbifold
fixed points \cite{Georgi:2001ks}; these have been considered
recently from a somewhat different perspective in
refs.~\cite{Dvali:2000hr, Dvali:2001gm, Carena:2001xy}. The
localized kinetic terms have the form:
\beq
\label{localizedkinetic}
\int d^5 x~ c ~\delta(y)~ \partial_\mu
\Phi
\partial^\mu \Phi
\eeq
and   introduce a new length scale, $c$, in the problem. The
localized kinetic terms can significantly change the spectrum of
the nonzero modes \cite{Dvali:2000hr}, and hence the
$\phi$-dependence of the Casimir energy. We calculate, in Section
\ref{branecasimir}, the Casimir energy in the theory with the
inclusion of such terms and find that it is possible to achieve a
minimum for $\phi$ with vanishing cosmological constant; we also
discuss localized mass terms.

Finally, even though we will not pursue it here, we should mention
the possibility to combine the ``minimal" and ``nonminimal"
approaches, exploring the fact that the compact space is an
orbifold and at the same time introducing new fields for the
purpose of size stabilization. This is much in the spirit of
mechanisms considered before in codimension one and two in
refs.~\cite{Goldberger:1999uk, Arkani-Hamed:2000dz}. These are
classical mechanisms: one postulates that the fixed points are
sources for some bulk field, e.g., a scalar field. In codimension
two, see \cite{Arkani-Hamed:2000dz}, one introduces two kinds of
massless fields, ``dual" to each other, and finds that the energy
of one type grows with size while that of the other decreases (to
achieve that, one has to impose certain boundary conditions at the
fixed points), leading to a minimum for the size. Other
``nonminimal" mechanisms have been discussed in \cite{Luty:2000ec,
Luty:2000cz}. We note that if one does not want to achieve a
hierarchically large radius, it is possible to use a classical
mechanism of the type discussed above to balance a repulsive
Casimir force.

\subsection{Gravity and massless matter fields.}
\label{5dcasimirmassless}

In this Section, we calculate the Casimir energy due to the
gravitational field and to other massless fields, with both
periodic and antiperiodic boundary conditions on the $S^1$. Here we
assume that the contributions of the various terms at the fixed
points can be neglected, while  Section \ref{branecasimir} contains
a discussion  of the more general case.

Most of the discussion in this Section is not new. For
completeness, we briefly review the calculation of
ref.~\cite{Appelquist:1983vs}  of the contribution to the Casimir
energy of the fields $g_{\mu\nu},A_\mu,\phi$. We then generalize
the calculation to include matter fields with both periodic and
antiperiodic boundary conditions on the $S^1$. Finally, we give the
generalization to the orbifold case. We will see that, if the
contributions from fixed-point-localized terms are negligible,
  the Casimir energy of the $S^1/Z_2$ orbifold is one half
 that of the $S^1$ compactification.

Later on, in Section \ref{applications}, we   apply the formulae
obtained for various fields to calculate the potential for $\phi$
generated in the models of ref.~\cite{Barbieri:2001vh}, where the
supersymmetric standard model lives in the five dimensional bulk.
We show that in these models the Casimir potential for $\phi$ is
repulsive, hence (for vanishing coefficients of the counterterms)
the compact space tends to expand to infinite size.

We begin with the Casimir energy of the gravity sector. As in
ref.~\cite{Appelquist:1983vs}, only the nonzero modes contribute to
the potential for $\phi$, and the contribution of the gravitational
multiplet equals that of five massless real scalar fields. This can
be seen after appropriate gauge fixing and is essentially due to
the fact that the five-dimensional graviton has five polarization
states  \cite{Appelquist:1983vs}. Thus, it is enough to compute the
vacuum energy of a single massless scalar field,
 with periodic boundary conditions along the $S^1$:
 \beqa
 \label{masslessscalarperiodic}
 V^{+,scalar} ~&=&~ {1 \over 2} \sum\limits_{n = - \infty}^{\infty}
 \int {d^4 k\over (2 \pi)^4} \log\left( k^2 + {\pi^2 n^2 \over
 \phi~ L^2}\right) \nonumber \\
 & \equiv & - {d \over d s} \zeta^{+,scalar} (s)\big\vert_{s = 0}~.
 \eeqa
The second line indicates that we use $\zeta$-function
regularization to calculate the Casimir energy. Infinite
contributions to the Casimir energy  are thrown out by the
regularization and counterterms have to be added by hand, as
discussed above. The periodic scalar $\zeta$-function is defined
as:
\beqa
\label{masslessscalarzeta}
\zeta^{+,scalar} (s)
&=&~ {1 \over 2} \sum\limits_{n = - \infty}^{\infty}
 \int {d^4 k\over (2 \pi)^4}  \left( k^2 + {\pi^2 n^2 \over
 \phi~ L^2} \right)^{-s} \nonumber \\
 & = & {1\over 16 \pi^2} {1\over (2-s)(1-s)}{\pi^{4 - 2s} \over
 L^{4 - 2s} \phi^{2-s}} ~\zeta(2 s - 4)~,
 \eeqa
 with $\zeta(s)$---the Riemann zeta function. Plugging
 into eqn.~(\ref{masslessscalarperiodic}) yields for the massless periodic
 scalar contribution to the potential in eqn.~(\ref{5dreduction}):
 \beq
 \label{vscalarperiodic}
V^{+,scalar}(\phi) ~=~{\pi^2 \over 16} ~{- \zeta^\prime(-4) \over
\phi^2 L^4}~
=~ - {3 \zeta(5) \over 64 \pi^2} ~{1 \over \phi^2 L^4}~,
\eeq
which is the result of \cite{Appelquist:1983vs}. We thus find that
the contribution of the gravitational fluctuations to the Casimir
energy, equal to $5 V^{+,scalar}(\phi)$, is attractive
($\zeta(5)\simeq 1.034$), i.e. the circle tends to shrink to zero
size.

Having in our disposal the result for the periodic massless scalar,
we can easily enumerate the Casimir contributions of all other
periodic massless fields, e.g. by using knowledge about
five-dimensional supersymmetry multiplets. Below, we summarize the
results for all massless periodic fields of interest:
\beqa
\label{allfieldsperiodic}
V^{+,graviton}(\phi) &=& 5 ~V^{+,scalar}(\phi) ~, \nonumber \\
 V^{+,fermion}(\phi) &=& - 4~ V^{+,scalar}(\phi) ~, \nonumber \\
V^{+,vector}(\phi) &=&  3 ~V^{+,scalar}(\phi)  ~, \\
V^{+,gravitino}(\phi) &=& - 8 ~V^{+,scalar}(\phi) ~. \nonumber
\eeqa

It is also of interest to compute the contribution of matter fields
(hypermultiplets) with antiperiodic boundary conditions on the
$S^1$. The scalar contribution $V^{-,scalar}(\phi)$ is given by the
same expression as (\ref{masslessscalarperiodic}) except for the
sum being over half-integers. Thus, the formula for the
antiperiodic zeta function (\ref{masslessscalarzeta}) remains the
same safe for the replacement $\zeta(2 s - 4) \rightarrow \zeta(2 s
- 4, 1/2)$, where $\zeta(s,1/2) = (2^s - 1) \zeta(s)$ is the
generalized zeta function. Evaluating the derivative, we obtain a
repulsive potential for $\phi$ from an antiperiodic real scalar
field:
\beq
\label{vscalarantiperiodic}
V^{-,scalar}(\phi) ~=~ -  {15\over 16} ~ V^{+, scalar} (\phi)~,
\eeq
while the contribution of an antiperiodic fermion is attractive:
\beq
\label{vfermionantiperiodic}
V^{-,fermion}(\phi) ~=~ {15\over 4}~V^{+, scalar} (\phi)~.
\eeq

Finally, we come to the $S^1/Z_2$ orbifold Casimir energy.
Neglecting the effects of terms localized at the fixed points, the
orbifold amounts to simply throwing out all modes even under $y
\rightarrow -y$ (for odd fields) or odd (for even fields). Since
only the non-zero modes contribute to the Casimir energy, it is
easy to see that in each case this amounts to throwing out half the
modes in the sum over $n$ in (\ref{masslessscalarperiodic}). (We
note that an additional factor of $1/2$ will occur because in the
orbifold we only integrate from $y=0$ to $y=L$. This, however, will
be irrelevant for us.)

As noted in the Introduction, including the contribution to the
Casimir energy of the massless fields alone (without including
brane tensions) it is not possible to obtain a minimum for $\phi$
with vanishing cosmological constant. We will, in what follows, use
the results from this Section to discuss ways of obtaining a
minimum when other possible contributions are also included.

\subsection{Casimir energy and radius stabilization with massive matter fields.}
\label{5dcasimirmassive}

In this Section, we calculate the contribution of both periodic and
antiperiodic massive fields to the Casimir energy. We begin with
the periodic case, when the vacuum energy for a real scalar of mass
$\mu$ is given by:
\beq
\label{massivescalarperiodic}
V^{+,scalar} (\phi, \mu)~= ~ {1 \over 2} \sum\limits_{n = -
\infty}^{\infty}
 \int {d^4 k\over (2 \pi)^4} \log\left( k^2 + {\pi^2 n^2 \over
 \phi~ L^2} + {\mu^2\over \phi^{1/3}} \right) \nonumber \\
  \equiv - {d \over d s} \zeta^{+,scalar}_\mu (s)\big\vert_{s = 0}~.
 \eeq
  The massive periodic scalar $\zeta$-function is:
  \beqa
\label{massivescalarzeta}
\zeta^{+,scalar}_\mu (s)
&=&~ {1 \over 2} \sum\limits_{n = - \infty}^{\infty}
 \int {d^4 k\over (2 \pi)^4}  \left( k^2 + {\pi^2 n^2 \over
 \phi~ L^2} + {\mu^2\over \phi^{1/3}} \right)^{-s} \nonumber \\
 & = & {1\over 32 \pi^2} {1\over (2-s)(1-s)}{\pi^{4 - 2s} \over
 L^{4 - 2s} \phi^{2-s}} ~F(s-2;0, \mu L \phi^{1\over3}/\pi)~,
 \eeqa
 where $F(s;0,c)$ is the series:
 \beq
 \label{epstein}
 F(s;0,c)~
 =~\sum\limits_{n = - \infty}^{\infty} \left(n^2 + c^2 \right)^{-s}~,
 \eeq
 and can be evaluated as described in the Appendix. The
 contribution of an antiperiodic scalar field can also be evaluated---the
 only change is that the sum in (\ref{epstein}) is over half integers.

 The results for the Casimir
 contribution $V^{+,scalar} (\phi, \mu)$, $V^{-,scalar} (\phi, \mu)$
 of periodic or antiperiodic massive scalars
 respectively---omitting both finite
 and infinite  contributions to the
bulk cosmological constant---can be written in a common form:
 \beqa
 \label{massiveperiodicscalarpotential}
 && V^{\pm,scalar} (\phi, \mu) = - {3 \over 64 \pi^2 }
 {1 \over L^4 \phi^2} \times  \nonumber \\
 &&\times \left[
  {\rm Li}_5 \left(\pm e^{- 2 L \mu \phi^{1\over 3}} \right) +
2 L \mu \phi^{1\over 3} ~  {\rm Li}_4 \left( \pm e^{- 2 L
\mu \phi^{1\over 3}}\right) +
{4\over 3} L^2 \mu^2 \phi^{2\over 3}~   {\rm Li}_3
\left( \pm e^{- 2 L \mu \phi^{1\over 3}} \right)
  \right]~,
  \eeqa
  where ${\rm Li}_n(x)$ are the polylogarithm  functions.
   Since ${\rm Li}_5(1) = \zeta(5)$ and
   ${\rm Li}_5(-1)= - 15 \zeta(5)/16$,
  the above expression reduces to the massless
   formulae for the periodic (\ref{vscalarperiodic})  and
  antiperiodic (\ref{vscalarantiperiodic}) case
  in the small mass or small radius ($\phi \rightarrow 0$) limit.
As in the massless case, the contribution of a massive fermion on
the $S^1$ is equal to minus 4 times the real scalar contribution;
the $Z_2$ orbifold contribution is one-half the unorbifolded
contribution in each case.\footnote{The potential
eqn.(\ref{massiveperiodicscalarpotential}) was independently
derived in Ref. \cite{Delgado:1999qr}.}

The bulk mass $\mu$ introduces a new scale in the problem, leading
one to expect that stabilizing the radius using the Casimir energy
of massive fields should be possible in certain cases. To see what
the necessary conditions are, consider the following   toy model
with a single scale $\mu$. Let there be a periodic field (scalar or
fermion; or a number of fields) of mass $\mu$ and a massless sector
(consisting of gravity and possibly other massless fields).
Introducing the variable $x = 2  \mu L \phi^{1/3}$, we can write
the total potential for $x$ in the following form:
\beq
\label{massivepotential}
V(x) ~=~ {|const.| \over x^6}\left(\alpha x^5  + \beta \left( {\rm
Li}_5(e^{-x}) + x ~{\rm Li}_4(e^{-x})+ {x^2\over 3}~{\rm
Li}_3(e^{-x}) \right) + \gamma
\right)~\equiv~ x^{-6} f(x) ~,
\eeq
where the first term is the cosmological constant counterterm,
while the second and third terms are the contributions
 to the Casimir energy of the
masssive and massless fields, respectively. That the potential
$V(x)$ can have a stable minimum is easy to understand
qualitatively. A stable minimum can occur only\footnote{It is
important to note that the term  multiplying $\beta$ is a
monotonically decreasing function of $x$ equal to  $Li_5(1)$ at $x
= 0$.} if the massive contribution dominates near the origin (and
is repulsive), while the massless term (with opposite sign) takes
over at larger distances, where the massive contribution is
exponentially suppressed. The normalization conditions
(\ref{normalizationconditions}) become, for the potential $V(x)$
(\ref{massivepotential}):
\beq
\label{conditionsM}
f'(x) = 0,~~~ f(x) = 0,
\eeq
and can be  shown to have solutions provided the coefficients
$\gamma$ and  $\beta$ have opposite signs, while
 $\alpha$ and $\beta$ have the same sign; in addition, the solution of
(\ref{conditionsM}) is a minimum if $\gamma < 0$. One also requires
that $|\gamma| < {\rm Li}_5(1)  |\beta| $ so that the massive
repulsive contribution dominates near $x = 0$ ensuring the
existence of a minimum.

The conclusion one can draw from this simple one-scale model is
that it is possible to have a stable minimum with zero cosmological
constant, provided: $i$.) the massless contribution is attractive,
$ii.$) the massive contribution is repulsive and dominant at small
radii, and, $iii.$) the bulk cosmological constant is fine-tuned
(and positive, corresponding to $\alpha > 0$); as usual, there is
also a runaway minimum at $x \rightarrow \infty$.\footnote{A
similar conclusion can be drawn if one  considers instead an
antiperiodic massive field---once again one requires that its
contribution to the Casimir energy be repulsive (hence the massive
field is an antiperiodic scalar) and dominant at short distances in
order to achieve a minimum (in this case $\alpha
>0$ as well).} Barring unnaturally large coefficients, the
radius is stabilized at a size of order the inverse mass of the
field, which is the only length scale in our toy model.

Using the results of this Section, it is straightforward  to
include fields of different masses, as well as brane tension
counterterms in the  discussion of radion stabilization.

\section{Casimir energy contribution of brane-localized kinetic terms.}
\label{branecasimir}

In this Section, we  calculate the contribution to the Casimir
energy of a massless scalar field while taking into account the
presence of  kinetic terms localized at the fixed points $y = 0,L$.
These terms can significantly affect the Kaluza-Klein spectrum of
the field and change the Casimir energy. Readers interested only in
the answer can skip to the final result,
eqn.~(\ref{boundarycasimir1}), and the following discussion of the
radius dependence of the Casimir energy with boundary terms.

In the metric background (\ref{5dinterval}) with  $g_{\mu\nu} =
\eta_{\mu\nu}$, $A_\mu
= 0$, $\phi = const.$, we take the quadratic  action of a real
periodic scalar field $\Psi$ to be:\footnote{Derivatives w.r.t. $y$
will also generally appear. These require a more careful treatment
in the thin wall limit we are studying here. For simplicity, we
consider only the terms of eqn.~(\ref{scalarlocalizedaction}).
These should be sufficient to illustrate the effects of localized
terms on the Casimir energy.}
\beq
\label{scalarlocalizedaction}
{1 \over 2} \int d^4 x \int\limits_{-L}^L d y\left[ \partial_\mu
\Psi   \partial^\mu \Psi + {1\over \phi}
(\partial_y \Psi)^2 + \left( 2 c_0 \phi^{-1/3} \delta(y)
+ 2 c_L \phi^{-1/3} \delta(L - y)\right) \partial_\mu
\Psi \partial^\mu \Psi \right] ~,
\eeq
where $c_0$ and $c_L$ are the two length scales introduced by the
localized kinetic terms. Expanding the field $\Psi(x,y)$ into
Fourier components $\Psi_k(y)$ of four-dimensional mass
$\sqrt{k^2}$, we find the equation of motion:
\beq
\label{eqofmotion}
\left[ \partial_y^2 + \phi k^2 + (2 c_0 \phi^{-1/3} \delta(y)
+ 2 c_L \phi^{-1/3}
\delta(L - y)) \phi k^2 \right] ~\Psi_k (y)
= 0~.
\eeq
Now we introduce the variables $\tilde{k}^2 \equiv \phi k^2$ and
$\tilde{c}_{0,L}
\equiv \phi^{-1/3} c_{0,L}$, and solve  eqn.~(\ref{eqofmotion}) for even periodic fields
$\Psi(y + 2 L) = \Psi(y), \Psi(-y)
= \Psi(y)$. We find that the values of $\tilde{k}$ for which the solution
has the appropriate discontinuities to match the delta functions
are determined by the solutions of:
\beq
\label{eigenvalueequation}
{\rm tan}( \tilde{k} L ) = -{ \tilde{k} (\tilde{c}_0 +
\tilde{c}_L) \over  1 - \tilde{c}_0 \tilde{c}_L  \tilde{k}^2 }~.
\eeq
When $c_{0,L} \rightarrow 0$, the equation simply gives the masses
of the Kaluza-Klein modes $|k_n|  = |n| \pi/L$. It is clear that
 the solutions of eqn.~(\ref{eigenvalueequation}) are also labeled by an
 integer, and we shall denote them by $\tilde{k}_n$. Note also that
 if $c_{0,L}$ are positive, there are no negative $k^2$ solutions,
  i.e. no tachyons.

Expanding the field   $\Psi(x,y) = \sum\limits_{n = 0}^{\infty}
\varphi_n(x) \Psi_{k_n}(y)$ in terms of the (even) modes $\Psi_{k_n}(y)$,
we find that the quadratic part of the four dimensional theory of the
``Kaluza-Klein" modes $\varphi_n(x)$ is governed by the action:
\beq
\label{kkaction}
{1\over 2} \int d^4 x \sum\limits_{m,n = 0}^{\infty}
\left(\partial_\mu \varphi_n \partial^\mu \varphi_m +
 {\tilde{k}_n^2\over \phi} \varphi_n \varphi_m \right) \gamma_{mn}~.
\eeq
The normalization coefficients obey:
\beq
\label{gamma}
\gamma_{mn} = \int\limits_{-L}^{L} d y \Psi_{k_n}(y) \Psi_{k_m}(y)
\left(1 + 2 \tilde{c}_0 \delta(y) + 2 \tilde{c}_L \delta(y-L)\right)
 = 0 ~{\rm for}~ m\ne n~.
\eeq
The orthogonality of the Kaluza-Klein wavefunctions follows from
the ``Schr\" odinger"  equation (\ref{eigenvalueequation}). The
diagonal normalization coefficients $\gamma_{nn}$ depend on $n$,
however, in contrast to conventional Kaluza-Klein
compactifications; this is important when considering interactions
involving the Kaluza-Klein modes \cite{Dvali:2001gm,
Carena:2001xy}. The normalization coefficients drop out of the
Casimir energy, however.

The Casimir energy of the real scalar field, even under the $Z_2$
orbifold projection, can be then written as follows:
\beq
\label{boundarycasimir}
V^{+,scalar}(\phi; c_{0,L}) ~ = ~{1\over 2} \sum\limits_{n =0
}^\infty \int {d^4 p \over 2 \pi^4} \log \left( p^2 +
{\tilde{k}_n^2\over
\phi} \right)
~\equiv ~ - {d\over d s}~ \zeta_{c_{0,L}}(s)\bigg\vert_{s
\rightarrow 0}~.
\eeq
The $\zeta$-function is that of the operator in (\ref{eqofmotion})
and is  given by:
\beq
\label{zetaboundary}
\zeta_{c_{0,L}}(s) ~ = ~  {1\over 32 \pi^2} {1\over (2-s)(1-s)}
{1 \over L^{4 - 2s} \phi^{2-s}} ~F(2 s - 4)~, ~~{\rm where} ~~ F(s)
\equiv
\sum\limits_{n
= 1}^\infty {1
\over x_n^s}~,
\eeq
where the sum is over the nonnegative roots of
eqn.~(\ref{eigenvalueequation}), which written in terms of $x =
\tilde{k} L$, has the form ${\rm tan} x  =  - x a/(1 - b x^2)$. The
coefficients $a,b$ are expressed in terms of the ratios of the
length scales $c_{0,L}$ from (\ref{scalarlocalizedaction}) to the
physical size of the orbifold $l_{phys} = L \phi^{1/3}$ as follows:
\beqa
\label{aandb}
 a~ &=&~ {\tilde{c}_0   + \tilde{c}_L \over L}  ~=~ {c_0   +  c_L \over l_{phys}}~, \\
b ~&=&~ {\tilde{c_0} \tilde{c}_L \over L^2} = { c_0 c_L \over
l_{phys}^2} ~.\nonumber
\eeqa

The sum $F(s)$ of eqn.~(\ref{zetaboundary}) can be taken by means
of contour integrals as follows. Introduce the function:
\beq
\label{ef}
f(z)
= a z + (1
- b z^2) {\rm tan} z~,
\eeq
and consider the contour integral:
\beq
\label{integral}
I(s) ~=~ {1 \over 2 \pi i}~ \int\limits_C  d z ~ {1\over z^s}~
{f^\prime (z)
\over f(z)} ~,~~ C = C_\infty + C_+ + C_\epsilon + C_-~,
\eeq
where the contour $C_\infty$ is  an infinite semicircle in the
${\rm Re} z > 0$ half-plane, $C_+$ runs  along the imaginary axis
from $z
= i
\infty$ to $z
= i \epsilon$, $C_-$---from $z = - i \epsilon$ to $-i\infty$, and $C_\epsilon$ is
a small semicircle from $z = i \epsilon$ to $z = - i \epsilon$ in
the ${\rm Re} z > 0$ half-plane. The sum $F(s)$ in
eqn.~(\ref{zetaboundary}) can be written as:
\beq
\label{sum}
F(s) \equiv \sum\limits_{n >0} {1 \over x_n^s} = I(s) + {1 \over
\pi^s} \sum\limits_{n \ge 0}
{1 \over \left(n + {1\over2}\right)^s}~=~ I(s) + {2^{s} - 1 \over
\pi^s} \zeta(s)~,
\eeq
where we used the fact that $I(s)$ is determined by the poles of
$f^\prime/f$ inside the contour $C$, which occur at the zeros ($z =
x_n$) and poles ($z = (n +1/2) \pi$)  of  $f(z)$, eqn.~(\ref{ef}).

We are interested in computing  $F^\prime(-4)$, which, as usual is
defined  by analytic continuation. The integral over $C_\infty$
vanishes for a sufficiently large positive Re($s$). The rest of the
integral can be explicitly evaluated and written as follows:
\beqa
\label{integralexplicit}
I(s) &=& {s\over \pi} \sin \left({\pi s \over 2}\right)
\int\limits_\epsilon^\infty d y ~y^{-s - 1} \log  {a y + (1 + b y^2) {\rm tanh}y \over 1 + a y + b y^2}  \nonumber \\
&+&{s\over \pi} \sin\left({\pi s \over 2}\right)
\int\limits_\epsilon^\infty d y ~y^{-s - 1} \log(1 + a y + b y^2) \\
&-& {s \epsilon^{-s} \over 2 \pi }
\int\limits_{-\pi/2}^{\pi/2} d \theta~ e^{- i s \theta} ~ \log f(\epsilon e^{i\theta}) ~-~ {\epsilon^{-s}\over 2}
\cos \left({\pi s \over 2}\right)~ . \nonumber
\eeqa
It is easy to see from (\ref{integralexplicit}) that the
divergences as $\epsilon \rightarrow 0$ cancel. This form of the
integral is also appropriate for analytic continuation to negative
$s$. The first integral is well-behaved for $s < 0$. The second
integral can be explicitly evaluated in terms of hypergeometric
functions; its analytic continuation for negative $s$ vanishes as
 $\epsilon \rightarrow 0$. The analytic continuation of the last line for $s
<0$ also vanishes in this limit. Thus, for negative $s$, we only
have the first integral, where the limit $\epsilon \rightarrow 0$
can be taken safely and we arrive at our formula for the analytic
continuation of $F(s)$ to negative $s$:
\beq
\label{EFofs}
F(s) = {2^{s} - 1 \over \pi^s} \zeta(s) + {s\over \pi} \sin
\left({\pi s
\over 2}\right)
\int\limits_0^\infty d y ~y^{-s - 1} \log
  {a y + (1 + b y^2) {\rm tanh}y \over 1 + a y + b y^2}~,
\eeq
which gives rise to:
\beq
\label{derivative}
-{d \over d s} F(2 s - 4)\bigg\vert_{s = 0} ~=~
{15 ~\pi^4 \over 8} ~\zeta^\prime(-4) + 4 \int\limits_0^\infty d y
~y^3
\log
  {a y + (1 + b y^2) {\rm tanh}y \over 1 + a y + b y^2}~.
\eeq
In the limit $a = b = 0$, the expression (\ref{derivative}), upon
substitution in (\ref{boundarycasimir}), yields the Casimir energy
 (\ref{vscalarperiodic}) without the boundary terms.

The final result for the Casimir energy of the even periodic scalar
with boundary kinetic terms is, therefore:
\beq
\label{boundarycasimir1}
V^{+,scalar}(\phi; c_{0,L}) ~ =~{1 \over  64 \pi^2 L^4
\phi^2}  \left[ {15 \over 8} \pi^4  \zeta^\prime(-4) +
\rho\left({c_0 + c_L \over L \phi^{1/3}},
{c_0 c_L \over L^2 \phi^{2/3}}\right) \right]~,
\eeq
with $\rho(a,b) = 4 \int\limits_0^\infty d y y^3
\log\left[(a y + (1 + b y^2) {\rm tanh}y)/(  1 + a y + b y^2)\right]$.
While the form
of the $\phi$-dependence is rather complicated and  can be studied
in detail only numerically, eqn.~(\ref{boundarycasimir1}) can be
used to illustrate some qualitative features of the Casimir energy
with boundary terms.

Consider first the large-radius behavior---the limit $\phi
\rightarrow
\infty$. In this limit, the coefficients $a, b$ vanish and
eqn.~(\ref{boundarycasimir1}) reduces to
(\ref{vscalarperiodic})---except for a factor of 1/2 due to the
orbifold---giving thus an attractive potential at large $\phi$:
\beq
\label{asymptoticinfty}
 V^{+,scalar}(\phi \rightarrow \infty; c_{0,L})
  ~\sim~  - {3 \zeta(5) \over 128 \pi^2} ~{1 \over \phi^2 L^4} + \ldots~.
\eeq
Thus, as expected, the boundary terms do not affect the
large-radius behavior of the Casimir energy.

The limit of small radii $\phi \rightarrow 0$ is more involved.
 To study it, note that the function $\rho$ of (\ref{boundarycasimir1})
  can be written as:
  \beq
  \label{rho1}
\rho =
 4 \phi^{4/3} \int\limits_0^\infty d y y^3\log\left[{ y +
 (1 + \hat{b} y^2) {\rm tanh}\phi^{1/3} y \over
  1 + y + \hat{b} y^2}\right]~,
\eeq
 where
the hat  indicates that the $\phi$ dependence has been scaled out
of $b$; in addition, to simplify our formulae, we have chosen  the
arbitrary length scale $L$ such that $a \equiv 1$.

In the limit when  the boundary term at one of the fixed points
vanishes, i.e. $\hat{b} =0$, it is easy to see that $\rho
\rightarrow 0$ as $\phi
\rightarrow 0$ (the integral in (\ref{rho1})
diverges as $\phi \rightarrow 0$, but
more slowly than
 the $\phi^{4/3}$ prefactor, such that the whole
 expression (\ref{rho1}) vanishes).
 The small-$\phi$ potential is therefore repulsive.
  Thus, in the small radius limit
 the Casimir energy for the periodic scalar with
boundary term at only one of the fixed points is repulsive, while
the large-radius behavior, eqn.~(\ref{asymptoticinfty}), is
attractive. We   conclude that in the limiting case when the
boundary kinetic terms at one of the fixed points, say $c_L = 0$,
vanishes, the Casimir energy exhibits a minimum for $\phi$. The
minimum occurs for values of the physical radius of order the
length scale $c_{0}$ set by the nonvanishing boundary kinetic term.

 The behavior of the Casimir energy with one nonzero boundary term
 can be inferred already  from the eigenvalue equation
(\ref{eigenvalueequation}) in the limit when one of
 the boundary terms vanishes---then the lowest
eigenvalues are approximately the periodic Kaluza-Klein modes $n
\pi/L$, while the large eigenvalues are the antiperiodic ones, i.e.
$(n + 1/2) \pi/L$. One expects that the large-radius limit of the
Casimir energy is dominated by the lowest eigenvalues---hence the
attractive behavior characteristic of a periodic scalar---while the
small-radius limit depends on the large eigenvalues, leading one to
expect a repulsive behavior, as for an antiperiodic scalar.
 The behavior of the Casimir
energy (\ref{boundarycasimir}) in the two limits $\phi
\rightarrow \infty$ and $\phi \rightarrow 0$ confirms this expectation.

In the case where both boundary terms are present, the analysis
requires more care. Upon analyzing the integral (\ref{rho1}) one
finds that when $a \ne 0, b\ne 0$ the Casimir energy is attractive
at small $\phi$ as well (as in the case without boundary terms).
Thus, the existence or not of a minimum depends on relative
strength of $a$ and $b$. By comparing to the $a \ne 0$, $b=0$ case,
one expects that for a sufficiently small ratio of $b/a$, a local
minimum will still persist. A numerical analysis confirms this
expectation---it is sufficient to have $c_0/c_L \sim
.3$ in order for a (local) minimum to exist. There is no minimum if
$c_0 = c_L$, however.

Thus, we conclude that the radius can be stabilized by the boundary
kinetic terms, provided there is some asymmetry between the fixed
points.\footnote{We note that the classical mechanisms for radius
stabilization of  \cite{Goldberger:1999uk, Arkani-Hamed:2000dz},
also rely on an assymetry between the fixed points (``branes").} As
usual, the cosmological constant at the minimum can be adjusted to
zero by tuning the coefficients of the counterterms. Finally, while
for a periodic scalar with asymmetric boundary kinetic terms the
minimum is local, we expect that for a periodic fermion the
boundary kinetic terms  will yield a global minimum.

To conclude this Section, we consider the effect of brane kinetic
terms on fields that obey antiperiodic boundary conditions. Solving
the scalar equation (\ref{eqofmotion}) but imposing antiperiodic
boundary conditions:
\beq
\Psi_k(y+2L) = -\Psi_k(y)~,
\eeq
we find that the modes can be divided in two types:
\noindent i) modes that are even about $y = L$ with a spectrum
determined by:
\beq
\label{EvenAnti}
\tan(\tilde{k} L) = \frac{1}{\tilde{c}_L \tilde{k}}~,
\eeq
and \noindent ii) modes that are odd about $y = L$ with a spectrum
determined by:
\beq
\label{OddAnti}
\tan(\tilde{k} L) = \frac{1}{\tilde{c}_0 \tilde{k}}~.
\eeq
The Casimir energy of such modes is calculated in exactly the same
way as for eqns.~(\ref{boundarycasimir}) and (\ref{zetaboundary}).
In the present case, the appropriate analytical continuation,
analogous to eqn.~(\ref{derivative}), yields:
\beq
\label{derivativeAnti}
-{d \over d s} F(2 s - 4)\bigg\vert_{s = 0} ~=~
{15 ~\pi^4 \over 8} ~\zeta^\prime(-4) + 4 \int\limits_0^\infty d y
~y^3
\log
  {\tilde{c}_i L^{-1} y {\rm tanh}y +1\over \tilde{c}_i L^{-1} y +1}~.
\eeq
The zeros of the numerator inside the logarithm are those of
eqns.~(\ref{EvenAnti}) and (\ref{OddAnti}), while the first term
subtracts the contributions from the poles of the tangent.
Individually, each of these expressions produces a maximum for the
radion $\phi$ (remember that $c_i \sim \phi^{-1/3}$). This can be
understood directly from eqns.~(\ref{EvenAnti}) and
(\ref{OddAnti}). We note that as $\phi \rightarrow \infty$, the
spectrum is given by $\tilde{k} L = (n+\frac{1}{2})\pi$, \ie for
large radius the brane terms do not affect the spectrum of an
antiperiodic field. In particular, for large $\phi$ the potential
is repulsive. On the other end, for $\phi \rightarrow 0$ the
spectrum is given by $\tilde{k} L = n \pi$, \ie as for a field with
periodic boundary conditions, which produces an attractive
potential. Therefore, the potential must have a maximum at some
finite $\phi$.\footnote{This behavior can be confirmed numerically
from eqn.~(\ref{derivativeAnti}); note that the integrand in
(\ref{derivativeAnti}), after conveniently choosing $L = c_i$,
depends only on $\phi$.}

\subsection{Exact tree-level  propagator including brane-localized terms}
\label{branepropagator}

In the previous Section, we showed that brane-localized terms can
have an important effect on the Casimir energy and actually produce
a stable minimum for the radion potential. This is not completely
surprising since the coefficients of these terms introduce in
general a new length scale in the problem.  We showed that the
inclusion of brane kinetic terms can stabilize the radion at a
scale of the order of $c_{0,L}$ of
eqn.~(\ref{scalarlocalizedaction}).  Since these coefficients have
dimension of length, and since we would like to stabilize the
radius at a value somewhat larger than the inverse cutoff of the
theory, it follows that the coefficients of the brane-localized
kinetic terms need to be larger than the fundamental length scale
if they are to be relevant for radion stabilization. In particular,
their effects should be treated  exactly, as we did in the
calculation of the Casimir energy.\footnote{One way localized
kinetic terms could arise in, e.g., a string
 construction would be through the
expectation values of fields, confined to propagate to the fixed
points of the orbifolds (``twisted sector" fields); a  similar
mechanism has been exploited in a field theory context in
\cite{Dvali:2001gm}.}

Then the question arises as to whether it is consistent to allow
these terms to be large, while treating other effects
perturbatively.  In this Section, we derive the exact tree-level
propagator including brane kinetic terms and argue that
perturbation theory does not break down, even if the quadratic
brane operators have anomalously large coefficients.

To this end, consider a real scalar field in 5-d flat spacetime
with action:
\beq
\label{action3}
S = \frac{1}{2} \int d^4x \int_{-L}^{L} dy \left[\partial_M \Phi
\partial^M
\Phi + 2 c_0 \delta(y) \partial_\mu \Phi \partial^\mu \Phi + 2 c_L
\delta(y-L) \partial_\mu \Phi \partial^\mu \Phi \right].
\eeq
In this Section, $M = 0,1,2,3,y$, and $\mu = 0,1,2,3$. We assume
that the fifth dimension is compactified on $S^1$ and, to simplify
notation, neglect the dependence on $\phi$, which can trivially be
included.
 The propagator obeys:
%\footnote{In this section we assume a metric signature
%$(+, -, -, - ,-)$.}
\beq
\label{PropEqOne}
\left(\partial_y^2 + p^2 + 2 c_0 p^2 \delta(y) +
2 c_L p^2 \delta(y-L)\right) G(p;y,y') = \delta(y-y')
\eeq
where we Fourier transformed in the four noncompact coordinates.
Since $G(p;y,y')$ is periodic in $y$ with period $2L$, we can
write:
\beqa
G(p;y,y') & = & \frac{1}{2L} \sum_n e^{-i \frac{\pi n}{L} y}
G_n(p;y')~,\\
\delta(y-y') & = & \frac{1}{2L} \sum_n e^{-i \frac{\pi n}{L} (y-y')}. \nonumber
\eeqa
Replacing these expansions back in eqn.~(\ref{PropEqOne}), dividing
by $p^2 -
\left(\frac{\pi n}{L}\right)^2$ and summing
$\sum_n e^{-i \frac{\pi n}{L} y}$, we obtain:
\beq
\label{PropEqTwo}
G(p;y,y') + 2 c_0 p^2 G(p;0,y')B(p,y) +  2 c_1 p^2
G(p;L,y')B(p,y-L)
= B(p,y-y')
\eeq
where we defined:
\beq
\label{Bfunction}
B(p,y) = \frac{1}{2L} \sum_n \frac{e^{-i \frac{\pi n}{L} y}}{p^2
- \left(\frac{\pi n}{L}\right)^2} \;.
\eeq

Evaluating eqn.~(\ref{PropEqTwo}) at $y=0$ and $y=L$ gives two
equations, from which $G(p;0,y')$ and $G(p;L,y')$ can be found as a
function of $B(p,y)$. Eqn.~(\ref{PropEqTwo}) then gives
$G(p;y,y')$. So, it remains only to evaluate the sum in
(\ref{Bfunction}).  This can be done by methods similar to those
used to evaluate the sums for the Casimir energy (see Appendix) and
we shall not repeat them here. The result is:
\beq
\label{ExplicitBfunction}
B(p,y) = \frac{1}{2p} \csc(pL)\cos\left(\left(1-\frac{y}{L}+ 2
\left[\frac{y}{2L}\right]\right) p L\right)~,
\eeq
where $[x]$ denotes the integer closest to $x$, but smaller than
$x$.

The propagator can then be written as follows:
\beq
\label{decomposition}
G(p;y,y') = G_{+}(p;y,y')+G_{-}(p;y,y')
\eeq
where $G_{+}(p;y,y')$ is the propagator for the even modes of the
field and $G_{-}(p;y,y')$ is the propagator for the odd modes, that
is (denoting $\Phi_{\pm}(p,y) = (\Phi(p,y) \pm
\Phi(p,-y))/2$):
\beqa
\label{EvenOddProp}
G_{\pm}(p;y,y') &=& \langle \Phi_{\pm}(p,y)\Phi_{\pm}(-p,y')\rangle
\\
&=&{1\over 4}\left( G(p;y,y') \pm G(p;-y,y') \pm G(p;y,-y') +
G(p;y,y') \right)~.\nonumber
\eeqa

The final expression for the propagators of the even and odd modes
is easily obtained  from eqns.~(\ref{PropEqTwo}) and
(\ref{ExplicitBfunction}). In the region   $y, y'\in [0, L]$, the
propagator of the even modes of the field is:
\beq
\label{EvenProp}
G_{+}^{(1)}(p;y,y') ~=~ - ~{\left[\cos(p y_<) -   c_0 p
\sin(py_<) \right] \left[\cos(p (L-y_>))
- c_L p \sin(p (L-y_>))\right]\over 2 p  \left[(c_0 c_L p^2
- 1)
\sin(p L) - (c_{0} + c_{L}) p \cos(p L)\right]} ~,
\nonumber
\eeq
where $y_{<}$ ($y_{>}$) denotes the smaller (larger) of $y$, $y'$
and $p \equiv \sqrt{p^{2}}$. We observe that $G_{+}(p;y,y')$ has
poles whenever $(c_0 c_L p^2
- 1)
\sin(pL) - (c_{0} + c_{L}) p \cos(pL) = 0$. This agrees with our result from
Section \ref{branecasimir}, see eqn.~(\ref{eigenvalueequation}). In
the same region  $y, y'\in [0, L]$,
 the propagator for the odd modes has
the much simpler  form:
\beqa
\label{OddProp}
G_{-}^{(1)}(p;y,y') &=& -\frac{1}{2p}\csc(p L) \sin(py_<) \sin(p
(L-y_>))~.
\eeqa
 We note that  $G_{-}(p;y,y')$   has poles
at $p = n \pi/L$, which is consistent with the fact that the odd
modes vanish at the location of the branes and therefore do not
feel the localized kinetic terms.  The orbifold projection amounts
to keeping only $G_{+}(p;0,0)$ for even parity fields and
$G_{-}(p;0,0)$ for odd parity fields.

The propagators in the other regions can be easily found from the
reflection properties of $\Phi_{\pm}(p,y)$ in $y$ according to
eqn.~(\ref{EvenOddProp}). Explicitly, when $y, y'
\in [-L, 0]$,
\beq
G_{\pm}^{(2)}(p;y,y') = G_{\pm}^{(1)}(p;-y,-y'),
\eeq
while for  $y \in [0, L]$ and $y' \in [-L, 0]$ or v.v.,
\beq
G_{\pm}^{(3)}(p;y,y') = \pm G_{\pm}^{(1)}(p;|y|,|y'|)~.
\eeq
For $y,y'$ outside the  $[-L,L]$ interval, $G_{\pm}(p;y,y')$ are
extended periodically in both $y$ and $y'$.

We also note that $G_{+}^{(1)}(p;y,y') \rightarrow
G_{-}^{(1)}(p;y,y')$ as both $c_{0}, c_{L}\rightarrow\infty$.  It
is easy to check, that in this limit the full propagator
$G(p;y,y')$ is precisely the propagator with Dirichlet boundary
conditions. This can also be understood by considering the
eigenmodes of the scalar field equation and noting that both the
even and odd modes vanish at the branes when $c_{0},
c_{L}\rightarrow\infty$.  Thus, the limit of very large kinetic
terms reduces to a problem where the branes act as perfect mirrors.

We are now ready to address the main goal of this
Section---analyzing the divergences occurring in the perturbative
expansion when the quadratic brane-localized terms are large, while
interaction terms are considered as perturbations. The expressions
(\ref{EvenProp}) and (\ref{OddProp}) allow us to study the behavior
of the propagator as a function of the 4-d momentum $p$ and thus
analyze the divergences that will appear in loops with this
propagator.

We find, that if $y \neq y'$ the propagator vanishes exponentially
for large Euclidean momenta, $p \gg L^{-1}$.  This shows that the
possible divergences in the theory are local in $y$, as expected.
When $y = y'$ there is a difference whether we are sitting inside
the bulk or at one of the branes.  In the former case, $G(p;y,y)
\sim \frac{1}{2p}$ which corresponds to a 5-d behavior.  In the
latter case, when $y = y'$ at one of the branes, say $y= 0$,
$G(p;0,0)
= G_{+}(p;0,0)
\sim
\frac{1}{2 p + 2 c_0 p^2}$ with $c_L$ in place of $c_0$ when $y = y' =
L$.  This is precisely the behavior found in
ref.~\cite{Dvali:2000hr} in the case of gravity.\footnote{The
models of  \cite{Dvali:2000hr} do not assume that the extra
dimensions are compactified. Their point is that the above 4-d
behavior due to the brane kinetic terms can make the
compactification unnecessary. The fact that our result reduces
precisely to theirs in the large momentum region is clear, since we
do not expect the large energy behavior to be sensitive to what may
happen far away from the brane in question.} On the other end, for
$p \ll L^{-1}$, the propagator behaves like $G(p;y,y')
\sim ((2 c_0+ 2 c_L+L)p^2)^{-1}$, which shows the 4-d behavior expected
from the finite size of the extra dimension.

If we add interactions to the theory, we can proceed to calculate
Feynman diagrams with our exact tree-level propagator,
eqn.~(\ref{decomposition}), or the appropriate one in an orbifolded
theory.  We imagine that the theory is valid below some scale
$\Lambda$, and that the 4-d momentum integrals are cutoff at
$\Lambda$. We also assume that the radius has been stabilized at $L
\gg
\Lambda^{-1}$, so that there is a region of energies where the 5-d
effective theory is valid.  The asymptotic behavior of the
propagator for large Euclidean momenta, discussed in the previous
paragraph, then shows that, in the bulk, the possible divergences
of Feynman diagrams are the same as in an uncompactified 5-d
theory, independent of the size of the brane kinetic terms.  The
convergence properties of Feynman diagrams at the branes, however,
depend on the size of $c_0$, $c_L$.  For definiteness, let us
concentrate on the $y=0$ brane and write $c_0 = \hat{c}
\Lambda^{-1}$.  If the brane couplings are perturbative, \ie
$\hat{c} \lsim 1$, the high energy behavior of the propagator
($\sim \frac{1}{p}$) is as expected in a 5-d theory.  If, on the
other hand $\hat{c} \gg 1$, the convergence properties are
improved---they are as in a 4-d theory---since for $p \gg c_0^{-1}
= \Lambda/\hat{c}$, the propagator behaves like $\sim \frac{1}{p^2}$.
Thus, it is consistent to assume that $c_{i} \gg \Lambda^{-1}$, in
which case the brane kinetic terms can play a role in the
stabilization mechanism for the radion.  If one is not interested
in obtaining a large hierarchy, moderately large $c_{i}$ might be
enough.  We remark that, as shown in ref.~\cite{Georgi:2001ks}, the
presence of brane kinetic terms seems to be quite generic in
theories with orbifold fixed points.

To conclude this Section, we note that it is straightforward to
include bulk and brane mass terms as well. Consider adding to the
action of eqn.~(\ref{action3}) the terms:
\beq
\Delta S = - \frac{1}{2} \int d^4x \int_{-L}^{L} dy \left[M^2
+ 2 m_0 \delta(y) + 2 m_1 \delta(y-L)\right] \Phi^2.
\eeq
The analogs of eqns.~(\ref{EvenProp}) and (\ref{OddProp}) for the
propagator derived from $S + \Delta S$ are:
\beqa
G_{+}^{(1)}(p;y,y') &=& - ~{\left[\cos(\rho y_<) -   b_0
\rho^{-1} \sin(\rho y_<) \right] \left[\cos(\rho (L-y_>))
- b_L \rho^{-1} \sin(\rho (L-y_>))\right]\over 2 \rho  \left[(b_0 b_L \rho^{-2}
- 1)
\sin(\rho L) - (b_{0} + b_{L}) \rho^{-1} \cos(\rho L)\right]} ~,
\nonumber \\
G_{-}^{(1)}(p;y,y') &=& - ~\frac{1}{2\rho}\csc(\rho L) \sin(\rho
y_<)
\sin(\rho (L-y_>))
\eeqa
where now $b_i = c_i p^2 - m_i$ and $\rho = \sqrt{p^2 - M^2}$.
These terms also induce a nontrivial radion potential.

%-------------------------------------------------------------
\section{Casimir energies in six dimensions compactified on a torus}
%-------------------------------------------------------------
\label{6dcasimir}

Now we consider the issue of the Casimir energy and radius
stabilization by quantum effects in the richer case of a 6-d
spacetime.  We consider compactifying two of the dimensions on a
torus, where the calculations run in parallel with our 5-d
analysis. The toroidal compactification is the one employed in the
composite-Higgs models of ref.~\cite{Arkani-Hamed:2000hv}. The
detailed spectrum and current experimental constraints   on the
size of the $T^2/Z_2$ orbifold from precision electroweak
observables are discussed in ref.~\cite{Appelquist:2000nn}.
Additional motivation for considering orbifold compactifications of
this type is the observation \cite{Dobrescu:2001ae} that in 6-d
standard-model global anomaly cancellation places severe
constraints on the number of identical generations, requiring $n_g
= 0~ {\rm mod} ~3$.

The main difference from the 5-d Casimir energy is that there are
now three moduli, which would have to be stabilized.  As we will
see, the dependence of the Casimir energies on these moduli is
nontrivial. To define a torus, one specifies a lattice in the plane
and identifies points that differ by a lattice vector. The three
moduli can be identified with the lengths of the two lattice
vectors and the angle between them. Alternatively, we can use the
area of the torus and the modular parameter $\tau = \tau_1 + i
\tau_2$. We consider the background interval:
\beq
\label{background6D}
ds^2 = \mathcal{A}^{-1} ~\eta_{\mu \nu} dx^\mu dx^\nu +
\mathcal{A} ~\gamma_{ij} dy^i dy^j
\eeq
where $y^i \in [0,L]$ and the metric $\gamma_{ij}$ on the torus is:
\beq
\gamma_{ij} = \frac{1}{\tau_2}
\left( \begin{array}{cc} 1 & \tau_1 \\\tau_1 & |\tau|^{2} \end{array} \right).
\eeq
In this parameterization, $det(\gamma) = 1$,  so $\mathcal{A}$ is
indeed the (dimensionless) area of the torus.  Further, the
parameterization is such that for fluctuations of the metric with
$\eta_{\mu \nu}
\rightarrow g_{\mu
\nu}(x^\mu)$, $\mathcal{A}(x^\mu)$, $\tau(x^\mu)$ depending only on
the four noncompact coordinates, we have:
\beq
M_6^4 \int d^6x \sqrt{G} R(G) = M^2 \int d^4x \sqrt{g} \left[ R(g)+
\frac{g^{\mu\nu} \partial_\mu \tau  \partial_\nu \bar\tau }{2\tau_2^2} +
\frac{g^{\mu\nu} \partial_\mu \mathcal{A} \partial_\nu \mathcal{A}}
{\mathcal{A}^2} \right ]
\eeq
where $M^2 = L^2 M_6^4$\footnote{M is not yet the physical 4-d
Planck mass that characterizes the gravitational interactions. Once
$\mathcal{A}$ gets a vacuum expectation value, it will be
convenient to do a constant Weyl rescaling $\tilde{g}_{\mu\nu} =
\mathcal{A}^{-1} g_{\mu\nu}$. It is $\tilde{g}$ that couples to all
fields. As a result, it is the coordinate independent combination
$\mathcal{A} L^2$ that appears everywhere and the physical 4-d
Planck mass is $M_4^2 = (\mathcal{A}L^2) M_6^4$.}, $R$ is the Ricci
scalar evaluated with the metric indicated in parenthesis and
$\bar\tau\equiv \tau_1 - i \tau_2$. There are also two massless
spin-one fields with their corresponding Kaluza-Klein towers. These
modes also contribute to the total Casimir energy, but they do not
play a role in the calculation of the Casimir energy itself (see
below) so we have not written them explicitly. We will see below
that their zero modes are projected out in the orbifold.

As in the 5-d case, it is enough to calculate the Casimir energy of
a real scalar field in the desired background (see discussion of
gauge fixing below). Thus, we begin with the vacuum energy of a
single massless scalar field propagating in the background
(\ref{background6D}), with periodic boundary conditions:
\beqa
V^{+,scalar} & = & \frac{1}{2  }\sum_{m,n=-\infty}^{\infty}
                   \int \frac{d^4k}{(2 \pi)^4} \log\left(R^2 k^2 +
                   \mathcal{A}^{-2} \tau_2^{-1} |m - n \tau|^2\right) \nonumber \\
             & = & -\frac{d}{ds} \zeta^{+,scalar} (s) |_{s=0}~,
\eeqa
where $R \equiv L/(2 \pi)$. Here we used again $\zeta$-function
regularization to define this expression, where now:
\beqa
\label{sum6D}
\zeta^{+,scalar} (s) & = & \frac{1}{2 }\sum_{m,n}{'}
                           \int \frac{d^4k}{(2 \pi)^4} \left(R^2 k^2 +
                           \mathcal{A}^{-2} \tau_2^{-1} |m - n \tau|^2\right)^{-s} \nonumber \\
                     & = & \frac{\pi^2}{L^4} \frac{ \mathcal{A}^{2s - 4}}{(s-2)(s-1)}
                           \sum_{m,n}{'} \frac{\tau_2^{s-2}}{|m - n \tau|^{2s-4}}~,
\eeqa
and the prime indicates that the zero mode is to be excluded. It is
clear from the second line that  $\zeta^{+,scalar} (s)$ is modular
invariant, reflecting the fact that the Casimir energy is
independent of the discrete choice of lattice vectors defining the
torus (these choices are related by $SL(2,Z)$ transformations of
the modular parameter $\tau$, see, e.g.,
\cite{DiFrancesco:1997nk}). We defer the evaluation of the sum in
(\ref{sum6D}) to the Appendix. The result for the potential is:
\beq
\label{Vscalar6D}
V^{+,scalar} ~=~ - \frac{\pi^2}{2L^4} ~\mathcal{A}^{-4}
~f(\tau,\bar{\tau})~,
\eeq
with (taking $\tau_2 > 0$ without loss of generality):
\beqa
\label{Vtau}
f(\tau,\bar{\tau}) & = & \frac{8\pi}{945} ~ \tau_2^3+
\frac{3  \zeta(5) }{\pi^4}~{1\over  \tau_2^2}  + \nonumber
\\ & & \hspace*{.3in} \frac{4}{\pi^2} \sum_{m=1}^{\infty}
\left( m^2~
{\rm Li}_3(q^m) + {3 \over 2 \pi} ~{1\over \tau_2}~ m ~ {\rm
Li}_4(q^m) + {3\over 4 \pi^2} ~{1\over
\tau_2^2 }~ {\rm Li}_5(q^m) + {\rm h.c.}\right)~,
\eeqa
where ${\rm Li}_n(x)$ are the polylogarithm functions and $q =
e^{2\pi i(\tau_1 + i  \tau_2 )}$.

A few comments are due on the properties of the function $f(\tau,
\bar\tau)$ in (\ref{Vscalar6D}). As mentioned above,  $f(\tau,
\bar\tau)$ is invariant
 under $SL(2,Z)$ modular transformations ($\tau \rightarrow (a \tau + b)/(c \tau + d)$,
 where $a,b,c,d$ are integers, obeying $a d - b c = 1$) due to the fact that tori with
 modular parameters related by an $SL(2,Z)$ transformation are diffeomorphic.
  Distinct tori have modular parameters $\tau$ taking values in the fundamental
 region $|\tau| \ge  1$, $1/2 > \tau_1 \ge -1/2$, $\tau_2 > 0$. It is therefore enough to study
 the behavior of the Casimir energy (\ref{Vscalar6D}) in the fundamental region. The  points
 $\tau = i$ and $\tau = e^{2 \pi i/3}$ of the fundamental region
  are fixed points of the transformations  $\tau \rightarrow - 1/\tau$ and
  $\tau \rightarrow -1/(1 + \tau)$, respectively, and are thus extrema of $f(\tau, \bar\tau)$.
  Numerically we find that the point $\tau = i$, corresponding to a rectangular torus, is a saddle point,
 the point $\tau = e^{2 \pi i/3}$,  corresponding to a torus of angle 120 degrees, is a minimum,
and that there are no other extrema of $f(\tau, \bar\tau)$. The
function  $f(\tau,\bar\tau)$ is positive definite and unbounded as
$\tau_2
\rightarrow
\infty$ in the fundamental region. Therefore, regarding the dependence of the Casimir energy  on the
modular parameter $\tau$, for fixed area $\cal{A}$, we can conclude
the following: i.) if the Casimir energy is attractive, as for a
periodic scalar, eqn.~(\ref{Vscalar6D}), the energy is unbounded
below as $\tau_2
\rightarrow \infty$. ii.) if the Casimir energy is repulsive,
the torus with $\tau = e^{2 \pi i/3}$ minimizes the potential  for
a fixed area. For repulsive total Casimir energy, therefore,
stabilizing the area modulus leads to stabilization of the modular
parameter.

To calculate now the total contribution to the Casimir energy due
to the gravitational fluctuations, it is necessary to count the
number of physical degrees of freedom. We write the 6-d metric as
$G_{M N} = G^{(0)}_{M N}+h_{M N}$, where $G^{(0)}_{M N}$ is the
background defined in eqn.~(\ref{background6D}) and $h_{M N}$ are
the quantum fluctuations. To begin, note that in a flat background
it is always possible to fix a ``transverse-traceless" gauge:
\beqa
G^{(0)M K}\partial_K \bar{h}_{M N} = 0 \nonumber \\
\bar{h} = 0 \nonumber
\eeqa
where $\bar{h}_{M N} = h_{M N} - \frac{1}{2} G^{(0)}_{M N} h$ and
$\bar{h} = G^{(0)M N}\bar{h}_{M N}$. This gauge fixing conditions
eliminate 7 of the original 21 polarizations. The fluctuations then
satisfy the wave equation $G^{(0)K L}\partial_K\partial_L h_{M N} =
0$. However, there is still a residual gauge freedom parameterized
by $h'_{M N} = h_{M N} +
\partial_M \xi_N + \partial_N \xi_M$, where $G^{(0)K L}\partial_K\partial_L \xi_M = 0$.
Working in Fourier space (discrete along the two compactified
dimensions), it is easy to see that, for modes with a nonzero
momentum in the compact dimensions, there is enough freedom to set:
\beqa
h_{\mu 6}(x, y^1, y^2) &=& - h_{\mu 5}(x, y^1, y^2) \nonumber \\
h_{55}(x, y^1, y^2) &=& h_{66}(x, y^1, y^2) \;. \nonumber
\eeqa
So the massive modes (from a 4-d perspective) include a tower of
spin-2 fields (with 5 polarizations each), a tower of spin-1 fields
(with 3 polarizations each) and a tower of real scalars. All of
these give a contribution to the Casimir energy which is 9 times
that of eqn.~(\ref{Vscalar6D}).

The zero modes include (before orbifolding) the 4-d graviton, two
gauge fields, and the three moduli that define the size and shape
of the torus; as usual, the zero modes do not contribute to the
Casimir energy. We also quote the contributions of other relevant
bulk fields:
\beqa
V^{+,graviton} &=& ~9 ~V^{+,scalar} ~,\nonumber \\ V^{+,Weyl
fermion} &=& -4 ~V^{+,scalar} \nonumber~, \\ V^{+,vector} &=&~ 4~
V^{+,scalar}~,
\\
 V^{+,2 form} &=& ~3~ V^{+,scalar}~, \nonumber \\
V^{+,gravitino} &=& -12~ V^{+,scalar}~. \nonumber
\eeqa
The simplest way to arrive at these results is by knowing the 6-d
$(1,0)$ supersymmetry multiplets. The toroidal compactification
preserves supersymmetry, hence the Casimir energy should vanish
within supermultiplets. The second equation follows from the fact
that the $(1,0)$ hypermultiplet contains a complex Weyl fermion and
two complex scalars, while the third follows from the content of a
$(1,0)$ vector multiplet (a vector field and a Weyl fermion).  The
self-dual two form field contribution is denoted by $V^{+, 2 form}$
(recall that a $(1,0)$ tensor multiplet consists of a Weyl fermion,
an anti-self-dual two form antisymmetric tensor field, and a real
scalar). Finally, the last line follows from the structure of the
$(1,0)$ gravity multiplet: graviton, gravitino, and  a self-dual
two form.

Note that if we orbifold the torus by a discrete subgroup of the
2-d rotation group, it is possible to ``freeze" the values of
$\tau_1$ and $\tau_2$ and concentrate on the $\mathcal{A}$
dependence of (\ref{Vscalar6D}) alone.\footnote{This would probably
be necessary in any case to obtain chiral fermions in the low
energy theory.} For example, if we write the coordinates in the
extra dimensions in complex notation as $z = y^1 + i y^2$ and
identify points $z \sim e^{2\pi i/n} z$, it is easy to see that
whenever $n > 2$, the requirement that the lattice that defines the
torus have the appropriate symmetry implies that $\tau_1$ and
$\tau_2$ must take on particular values.\footnote{In particular,
the case $n = 4$ corresponds to a square torus with $\tau_1 = 0$,
$\tau_2 = 1$.  In this case, the sum $\sum {'} (n^2+m^2)^{-s}$ in
eqn.~(\ref{sum6D}) can be evaluated more easily with the help of
the Jacobi identities.  We have checked that eqn.~(\ref{Vtau})
reproduces this result.}

The orbifolding will also project out some of the modes of the
gravity multiplet as can be seen by the requirement that the line
element be invariant. For example, the case $n=4$ corresponds to
the identifications $(y^{1},y^{2}) \sim (-y^{2},y^{1})$. Writing
the line element as:
\beq
ds^2 = G_{\mu \nu} dx^\mu dx^\nu + A_{\mu i} dx^{\mu}dy^{i} +
\gamma_{ij} dy^i dy^j \; ,
\eeq
determines the transformation properties of the various metric
components:
\beqa
G_{\mu \nu}(x^{\mu}, -y^{2},y^{1}) &=& G_{\mu \nu}(x^{\mu}, y^{1},
y^{2}) \nonumber \\ A_{\mu 1}(x^{\mu}, -y^{2},y^{1}) &=&
- A_{\mu2}(x^{\mu}, y^{1}, y^{2}) \nonumber \\ A_{\mu
2}(x^{\mu},-y^{2},y^{1}) &=& A_{\mu 1}(x^{\mu}, y^{1}, y^{2}) \\
\gamma_{11}(x^{\mu}, -y^{2},y^{1}) &=& \gamma_{22}(x^{\mu}, y^{1}, y^{2}) \nonumber \\
\gamma_{12}(x^{\mu}, -y^{2},y^{1}) &=& -\gamma_{21}(x^{\mu}, y^{1}, y^{2}) \nonumber
\eeqa
These parity assignments imply $A_{\mu i}(x^{\mu}, -y^{1},-y^{2}) =
-A_{\mu i}(x^{\mu}, y^{1},y^{2})$ and $\gamma_{12}(x^{\mu},
-y^{1},-y^{2}) = -\gamma_{12}(x^{\mu}, y^{1},y^{2})$. So the zero modes
of these components are projected out. Also all their Kaluza-Klein
components are forced to vanish at the positions of the orbifold
fixed points $(y^{1},y^{2}) = (0,0), (0,L), (L,0)$ and $(L,L)$. In
addition, the other ($y$-independent) zero modes  (apart from the
4-d graviton) satisfy $\gamma_{11}(x^{\mu})= \gamma_{22}(x^{\mu})$,
so there is only one modulus left, which can be identified with the
area of the torus.

Quite generally, there will be a discrete set of points that are
left invariant by the orbifold identification.  As discussed in the
previous sections, we expect the presence of terms in the action
localized on these fixed points.  Here we consider only the effect
of the brane tensions, which gives a contribution to the potential
for the area modulus $\mathcal{A}$:
\beq
\label{f1}
 T  \int d^6 x ~\delta(\vec{y})~\sqrt{\tilde{g}}  =
 T \int d^4 x ~\mathcal{A}^{-2}
\eeq
where $\tilde{g}$ is the induced metric at the relevant fixed point
and $T$---the corresponding brane tension. Adding these terms, as
well as a bulk cosmological constant:
\beq
\label{f2}
\Lambda  \int d^6 x ~\sqrt{G} =  \Lambda  L^2 \int d^4 x~ \mathcal{A}^{-1}~ ,
\eeq
to the contributions of the  massless fields (\ref{Vscalar6D}) it
is possible to obtain a minimum for $\mathcal{A}$ that satisfies
the normalization conditions (\ref{normalizationconditions}), as in
the 5-d case.

If we sum the contributions of all fields to the Casimir energy in
the model  of ref.~\cite{Arkani-Hamed:2000hv}, we find that the
potential for the area modulus is repulsive. We note that
ref.~\cite{Arkani-Hamed:2000hv} used a $Z_2$ orbifold, in which
case  there are two noncompact moduli; it is, however, possible to
generalize the construction to higher $Z_n$'s, affecting only the
massive spectrum, and thus project all but the area modulus
(alternatively, since the Casimir energy in the model with minimal
matter content is repulsive, as remarked after
eqn.~(\ref{Vscalar6D}), for fixed area the torus with $\tau = e^{2
\pi i/3}$ has minimal Casimir energy). With
just the area modulus, a minimum with zero cosmological constant
can be obtained after fine tuning the coefficients of the
counterterms (\ref{f1}), (\ref{f2}).

%-------------------------------------------------------------
\section{Applications}
%-------------------------------------------------------------
\label{applications}

In this Section, we consider the question of radius stabilization
in the supersymmetric 5-d model of \cite{Barbieri:2001vh} (see also
\cite{Arkani-Hamed:2000hv}).  This model is highly predictive, and
the authors of ref.~\cite{Barbieri:2001vh} were able to determine
not only the Higgs and superpartner masses, but also the
compactification radius of the fifth dimension, which, in fact,
sets the scale for all these masses.  However, they did not provide
a stabilization mechanism for the radius.  We are interested in
determining whether the radius can be stabilized by quantum effects
due to the original (minimal) field content.

Let us briefly review the main elements of the model of
ref.~\cite{Barbieri:2001vh}. One starts from a five dimensional
setup where the standard model superfields propagate in the bulk.
The fifth dimension is compactified on $S^{1}$, which is then
orbifolded by identifying points that are related by two $Z_{2}$
parities. The first one gives just the $S^{1}/Z_{2}$ orbifold
studied in Section \ref{5dcasimir} and breaks half the
supersymmetry. The second identification breaks the rest of the
supersymmetry giving precisely the standard model at low energies
(below the compactification scale).

The standard model gauge fields are described by 5-d vector
supermultiplets $(A^M, \lambda, \lambda', \sigma)$, which from a
4-d point of view decompose into one $N=1$ vector supermultiplet
$V=(A^\mu, \lambda)$ and one chiral superfield in the adjoint
representation $\Sigma(\phi_\Sigma, \psi_\Sigma)$, where
$\Phi_\Sigma
= (\sigma+i A^5)/\sqrt{2}$.  The standard model matter and Higgs fields
are described by hypermultiplets $(\Psi, \Phi, \Phi')_X$, where
$X=M$ ($=Q,U,D,L,E$) for matter fields and $X=H$ for the single
Higgs field of the model.  Here $\Psi$ is a Dirac fermion and
$\Phi$, $\Phi'$ are two complex scalars.  In 4-d language, these
correspond to two $N=1$ chiral supermultiplets $X(\Phi_X, \psi_X)$
and $X^c(\Phi^c_X,
\psi^c_X)$, where $\Psi = (\psi_X, \psi^c_X)$ and $\Phi' =\Phi^{c
\dagger}$.

In this model, matter and Higgs superfields are distinguished by
different parity assignments under the $Z_2 \times Z'_2$
projection. An equivalent way to describe this projection, which is
more suited to the notation employed in Section
\ref{5dcasimirmassless} is the following: we start from the
$S^1/Z_2$ orbifold discussed in Section \ref{5dcasimir} and break
the remaining 4-d $N=1$ supersymmetry by the Scherk-Schwarz
mechanism, imposing
\beq
\label{ScherkSchwarz}
\Phi(y+L) = R_{P} \Phi(y) \;,
\eeq
where $R_{P}$ is the R-parity of the unbroken supersymmetry of
$S^1/Z_2$ and $\Phi$ represents any component field.
Eqn.~(\ref{ScherkSchwarz}) is the same as imposing periodic
(antiperiodic) boundary conditions on the fundamental interval $[0,
L]$ on the fields with $R_{P}$ even (odd).  If we denote the
possible parity assignments by $(+,+)$, $(+,-)$, $(-,+)$, $(-,-)$
where the first entry refers to the $Z_{2}$ and the second to the
$R_{P}$ parity, the component field parity assignments are as given
in Table~1.
\begin{table}[t]
\label{parities2}
\centering
\begin{tabular}{c|c|c|c|c}
      & $(+,+)$   &  $(+,-)$  &     $(-,+)$           &    $(-,-)$           \\
\hline
$M$   & $\psi_M$  & $\Phi_M$  &  $\psi_M^{c\dagger}$  &  $\Phi_M^{c\dagger}$ \\
$H$   & $\Phi_H$  & $\psi_H$  &  $\Phi_H^{c\dagger}$  &  $\psi_H^{c\dagger}$ \\
$V$   & $A_\mu$   & $\lambda$ &  $\phi_\Sigma$        &  $\psi_\Sigma$       \\
\end{tabular}
\caption{Parity assignments for matter, Higgs and vector fields in the
model of ref.~\cite{Barbieri:2001vh}. The first assignment refers
to the $Z_{2}$ parity and the second to the $R_{P}$ parity of each
component field.}
\end{table}

In addition, one can write supersymmetric interactions at the
orbifold fixed points (using the notation of
ref.~\cite{Barbieri:2001vh}):
\beqa
\Delta L &=& \frac{1}{2}(\delta(y)+\delta(y-2L))\int
d^2\theta (\lambda_U Q U H) \nonumber \\ & & \mbox{} +
\frac{1}{2}(\delta(y-L)+\delta(y+L)) \int d^2\theta' (\lambda_D Q' D'
H' + \lambda_E  L' E' H') + {\rm h.c.}~,
\eeqa
which give rise to standard model quark masses after electroweak
symmetry breaking.

Ref.~\cite{Barbieri:2001vh} calculated the one-loop effective
potential for the standard model Higgs taking into account the
contribution of the top Yukawa coupling (the other Yukawa couplings
are much smaller and are neglected).  We note that this is
precisely the Casimir energy due to the $Q$ and $U$ matter
superfields in the presence of brane masses ($=\lambda_U\langle
H\rangle$). Adding the tree-level potential (from the D-terms) to
this quantum contribution, the authors of
ref.~\cite{Barbieri:2001vh} found a Higgs potential of the form
(see eqns.~(15) and (31) of ref.~\cite{Barbieri:2001vh}):
\beq
\label{VBHN}
V_{BHN}(\phi, H) = \frac{1}{\phi^{2} L^4} \hat{V}(\phi^{1/3} L H)~,
\eeq
where $\phi$ is the radion field in the parameterization
eqn.~(\ref{5dinterval}).  The minimization of this potential with
respect to $H$ fixes the combination $x \equiv \phi^{1/3} L H$ at a
value $x_{0}$ that can be determined numerically.  Knowledge of the
Higgs vev then determines the size of the extra space $\phi^{1/3}
L$. However, eqn.(\ref{VBHN}) is only part of the contribution to
the potential for $\phi$.  As discussed in Section \ref{5dcasimir},
all other fields, as well as possible cosmological and brane
tensions contribute to $V(\phi)$.

To analyze this potential, we first neglect the possible
contributions from brane kinetic terms. Then the spectrum for each
of the parity assignments is:
\beqa
\label{spectra}
m^{(+,+)}_n &=& 2n\pi/L      \nonumber \\
m^{(+,-)}_n &=& (2n+1)\pi/L  \nonumber \\
m^{(-,+)}_n &=& (2n+2)\pi/L            \\
m^{(-,-)}_n &=& (2n+1)\pi/L  \nonumber
\eeqa
with $n=0,1,2, \ldots$. We can obtain the Casimir energies in the
present orbifold from eqn.~(\ref{masslessscalarzeta}) by making the
replacement $L \rightarrow L/2$ and including a factor of
$\frac{1}{2}$ due to the first $Z_2$ orbifold, thus giving a total
factor of 8. Thus, the contribution from the KK tower of each real
degree of freedom is:
\beqa
\label{Potperdof}
V^{(+,+),real scalar}(\phi) &=& 8 V^{+,scalar}(\phi) \nonumber \\
V^{(+,-),real scalar}(\phi) &=& 8 V^{-,scalar}(\phi) \nonumber \\
V^{(-,+),real scalar}(\phi) &=& 8 V^{+,scalar}(\phi)  \\
V^{(-,-),real scalar}(\phi) &=& 8 V^{-,scalar}(\phi)~, \nonumber
\eeqa
where $V^{+,scalar}(\phi)$ and $V^{-,scalar}(\phi)$ are given in
eqns.~(\ref{vscalarperiodic}) and (\ref{vscalarantiperiodic}),
respectively. Taking into account the number of degrees of freedom
and parity assignments given in Table~1, we get for matter, Higgs
and vector supermultiplets the following contributions to the
scalar potential:
\beqa
\label{VMultiplets}
V^{(M)}(\phi) &=& -62 V^{+,scalar}(\phi) \nonumber \\
V^{(H)}(\phi) &=& +62 V^{+,scalar}(\phi)           \\
V^{(V)}(\phi) &=& +62 V^{+,scalar}(\phi) \nonumber
\eeqa
If we consider now the field content of the present model, namely
the $SU(3) \times SU(2)_{L} \times U(1)_{Y}$ vector multiplets,
three generations of matter hypermultiplets (less the third
generation $Q$ and $U$ superfields which were included in
eqn.(\ref{VBHN})), and one Higgs hypermultiplet, we get a total
contribution:
\beq
\label{massless}
 V^{massless}(\phi) = -1364 V^{+,scalar}(\phi) = +\frac{1023
\zeta(5)}{16}
\frac{1}{\phi^{2} L^{4}} \;.
\eeq
Thus the total radion potential is:
\beq
L^{4} V(\phi) = \alpha \phi^{-1/3} + \beta \phi^{-2/3} + \gamma
\phi^{-2}~,
\eeq
where $\gamma \equiv \hat{V}(x) + \frac{1023 \zeta(5)}{16}$
includes the contributions from all matter fields and we also
included a bulk cosmological constant $\alpha$, and possible brane
tensions $\beta$.  As we said before, the minimization with respect
to $H$, $\hat{V}'(x) = 0$, fixes $x = \phi^{1/3} L H$ at some
$x_{0}$ and also the value of $\hat{V}_{0} \equiv \hat{V}(x_{0})$
in eqn.(\ref{VBHN}) (which turns out to be positive).  The equation
of motion for $\phi$ then requires:
\beq
\label{minimum}
- \phi^{-1}\left( \frac{\alpha}{3} \phi^{-1/3} +
\frac{2\beta}{3} \phi^{-2/3} + 2 \gamma \phi^{-2}\right) = 0 \;.
\eeq
In addition, the requirement that the cosmological constant
vanishes implies:
\beq
\label{zerocosmo}
\alpha \phi^{-1/3} + \beta \phi^{-2/3} + \gamma \phi^{-2} = 0 \;.
\eeq
We note that solving eqns.(\ref{minimum}) and (\ref{zerocosmo})
imposes a fine-tuning condition among $\alpha$, $\beta$ and
$\gamma$. This is of course the cosmological constant problem,
which we are not trying to address here.  Note also that we found
that $\gamma > 0$, which is a consequence of the fact that the
matter content---which produces a repulsive potential---dominates
over the gauge and Higgs contributions.\footnote{And over the
gravity multiplet---it is easy to check that its attractive
contribution, neglected in (\ref{massless}), is much smaller than
the matter fields' repulsion.} We find then, that
eqns.~(\ref{minimum}) and (\ref{zerocosmo}) have a solution---which
is a minimum---provided $\alpha > 0$ and $\beta <0$ (and the
fine-tuning condition is satisfied). Thus, this mechanism could
only stabilize the radius if the bulk cosmological constant was
positive (de Sitter) and there are negative tension branes. The
first condition seems to be especially problematic, since a
positive cosmological constant seems to be incompatible with
supersymmetry. We also note that the minimum so obtained is not the
only minimum. Since the potential vanishes for large radius, there
is a second degenerate minimum at $\phi = \infty$.

It is  possible---and we leave this for future\footnote{It would be
also interesting to see to what extent  the predictions of the
model depend on the  values of the brane kinetic terms. In
particular, bounds on the coefficients of the kinetic terms from
single Kaluza-Klein-mode  production should be revisited---as
follows from eqns.~(\ref{kkaction},\ref{gamma}), the interactions
of these modes depend nontrivially on their excitation number.}
study---that the contribution from brane kinetic terms that we
discussed in Section \ref{branecasimir} could change the potential
enough to produce a minimum with zero 4-d cosmological constant
even when the bulk cosmological constant is negative. However, we
note that this would be at best a local minimum, and the true
minimum would be anti-de Sitter. To see this, we note that the
cosmological constant term always dominates at large separation of
the branes (i.e., at large $\phi$), so for negative bulk
cosmological constant the potential will always tend to zero from
below. If there is a minimum at a finite $\phi$ where the potential
vanishes, there will always be a second minimum where the potential
is negative. In this case one would have to worry about the
tunneling probability to the true AdS vacuum (recall, however, the
suppression of  tunneling to AdS space \cite{Coleman:1980aw}).

To conclude this Section, an issue which deserves a comment is that
of the scale of 5-d gravity (and of 6-d gravity from the previous
Section) in this model and the consistency of our approximations,
which ignored warping due to brane tension and bulk cosmological
constant counterterms. This is clearly a concern, since the 5-d
gauge theory is non-asymptotically free and the couplings blow up a
decade or so above the compactification radius, demanding a
transition to a  more fundamental description of the theory
somewhere in the (multi)-TeV region. If the fundamental scale of
5-d (or 6-d) gravity was in that range as well, we would need a
mechanism explaining the smallness of the observed gravitational
interaction in four dimensions.

One possibility in the present context is to assume that the higher
dimensional gravity scale is in the multi-TeV range while the
weakness of 4-d gravity is due to the presence of additional large
dimensions (e.g., of millimeter size) accessible only to gravity,
in the spirit of \cite{Arkani-Hamed:1998rs}. The radius
stabilization mechanism is then more complicated due to the
presence of these extra dimensions (also, some of our formulae for
the Casimir energies would need to be modified). More of a concern
in this regard is the neglect of the backreaction of brane tensions
and cosmological constant counterterms---since their size is
determined essentially by the inverse compactification scale (of
order TeV), which is not much smaller than the scale of the higher
dimensional gravity.

Another possibility---which makes neglecting the backreaction of
branes and bulk more palatable---is to assume that the 5-d (or 6-d)
gravity scale $M$ is close to the four dimensional observed
$M_{Planck}$ and the two are related in the usual way, e.g.,
$M_{Planck}^2 \sim L_{phys} M^3$ for the 5-d case. From a
low-energy point of view this constitutes an unexplained
fine-tuning, but may have its origin, as recently pointed out in
\cite{Antoniadis:2001sw}, in  ``little string theory" (the term
first appeared in \cite{Losev:1998hx}; for a review, see
\cite{Aharony:2000ks}). A major assumption required in order to
embed the models discussed in this paper in this framework is that
there is a window of energies above the compactification scale
where the field theory description is still valid.

\section{Concluding remarks.}
\label{conclusions}

This paper was devoted to the Casimir energy in five and six
dimensional field theory orbifolds and its effect on radion
stabilization. We were motivated by recent models of electroweak
symmetry breaking, which used 5-d (supersymmetric)
\cite{Barbieri:2001vh} and 6-d \cite{Arkani-Hamed:2000hv}
orbifolds.

We gave a general discussion of the divergences of the Casimir
energy and the counterterms required for their cancellation.
 We computed the Casimir energy for gravity and various massless fields,
obeying different boundary conditions. In the massive case, we
pointed out a  mechanism for stabilizing compact dimensions,
requiring that the massive contribution be repulsive, while the
massless be attractive.

We discussed in detail the influence of kinetic terms, localized at
the orbifold fixed points,  on the Casimir energy. We pointed out
that these brane-localized kinetic terms can also generate stable
minima for the radion. In the case of localized kinetic terms of
  size consistent with NDA \cite{Chacko:2000hg}, one can stabilize
  the compact dimension at a
size several times the fundamental length cutoff. For larger
brane-localized kinetic terms, that size can be larger. We argued
that the low-energy perturbative expansion in theories where brane
kinetic terms are larger than the inverse cutoff scale is
consistent (in accord with ref.~\cite{Dvali:2000hr}).

Applying our results to the 5-d and 6-d models of electroweak
breaking, mentioned above \cite{Barbieri:2001vh,
Arkani-Hamed:2000hv},  we found that, with the minimal field
content, the radion potential generated at one loop due to the
Casimir energy is repulsive in both cases. We found that it is
possible, by adding brane-tension and cosmological constant
counterterms to find a stable minimum for the radion.

However, in the supersymmetric 5-d case \cite{Barbieri:2001vh}, the
sign of the required bulk cosmological constant turned out to be
positive. Thus, deciding whether this fine-tuning is possible in
supergravity would need to wait for a full embedding of the model
in 5-d supergravity. Alternatively, our result might indicate that
a classical mechanism of stabilization is preferred; of course, it
would also have to be shown to be consistent with 5-d supergravity.

Finally, we note that, in this paper, we concentrated on static
issues, namely the existence of stable minima of the radion
field(s). Dynamical issues, such as the evolution of the universe
as a backreaction to the Casimir potentials are of interest as
well. It would also be extremely interesting to find what radius
stabilization mechanism is consistent with supergravity in the
supersymmetric case.

\section{Acknowledgments.}

We thank T. Appelquist, P. Horava, and M. Luty for useful
discussions.
  We
also acknowledge support of DOE contract DE-FG02-92ER-40704.

%-------------------------------------------------------------
\section{Appendix: Evaluation of Casimir sums}
%-------------------------------------------------------------
\label{appendix}

Here, we evaluate some of the sums that appear in the calculation
of the Casimir energies in 5- and 6-dimensional models, using
$\zeta$-function regularization.  Even though the methods for their
evaluation are standard, see \cite{DiFrancesco:1997nk, zeta}, we
show some of the details for completeness.  First, we evaluate the
single sum:
\beq
\label{SingleSum}
F(s;a,c) \equiv \sum\limits_{n=-\infty}^\infty\frac{1}{[(n + a)^{2}
+ c^{2}]^{s}}\;,
\eeq
which is convergent for sufficiently large positive Re($s$).  The
idea is to recast the sum into a form suitable for analytical
continuation to the values of $s$ of interest.  Setting $a=0$, this
is the sum that appears in the evaluation of the Casimir energy due
to a massive scalar field in 5-d, with periodic boundary
conditions. For antiperiodic boundary conditions we need $a =
\frac{1}{2}$. We start by noting that eqn.~(\ref{SingleSum})
defines a periodic function of $a$ with period 1.  So we can expand
eqn.~(\ref{SingleSum}) in Fourier series and write:
\beqa
F(s;a,c) &=& \sum_{p} e^{i 2 \pi p a} \int_{0}^{1}dy e^{-i 2 \pi p y}
\sum_{n}\frac{1}{[(n+y)^{2} + c^{2}]^{s}} \nonumber \\
&=& \sum_{p} e^{i 2 \pi p a} \sum_{n} \int_{n}^{n+1}dz e^{-i 2 \pi p z}
\frac{1}{[z^{2} + c^{2}]^{s}} \nonumber \\
&=& \sum_{p} e^{i 2 \pi p a} \int_{-\infty}^{\infty}dz e^{-i 2 \pi p z}
\frac{1}{\Gamma(s)} \int_{0}^{\infty}dt\,t^{s-1} e^{-(z^{2}+c^{2}) t}
\nonumber \\
&=& \frac{\sqrt{\pi}}{\Gamma(s)} |c|^{1-2s} \sum_{p} e^{i 2 \pi p a}
\int_{0}^{\infty}du\,u^{s-\frac{3}{2}} e^{-(u + \pi^{2} p^{2} c^{2} u^{-1})}
\nonumber \\
&=& \frac{\sqrt{\pi}}{\Gamma(s)} |c|^{1-2s}
\left( \int_{0}^{\infty} du\,u^{s-\frac{3}{2}} e^{-u} + 2 \sum_{p=1}^{\infty}
\cos(2\pi p a) \int_{0}^{\infty}du u^{s-\frac{3}{2}}
e^{-(u + \pi^{2} p^{2} c^{2} u^{-1})} \right) ~.\nonumber
\eeqa
In the third line we used the representation:
\beq
z^{-s} = \frac{1}{\Gamma(s)}\int_{0}^{\infty} dt\,t^{s-1}e^{-z t}~,
\eeq
while in the fourth line we performed the Gaussian integral over
$z$ and made the change of variable $u = c^{2}t$.  In the last
line, we recognize the integral representation of the modified
Bessel function:
\beq
K_{s}(|x|) = 2^{s-1} x^{-s} \int_{0}^{\infty}du\,u^{s-1}e^{-(u+\frac{x}{4u})}
\eeq
which is valid when $Re(x^{2}) > 0$. Thus,  we finally get for
(\ref{SingleSum}):
\beq
\label{FirstSum}
F(s;a,c) = \frac{\sqrt{\pi}}{\Gamma(s)} |c|^{1-2s}
\left(\Gamma \left(s-\frac{1}{2}\right) + 4 \sum_{p=1}^{\infty}
(\pi p |c|)^{s-\frac{1}{2}}\cos(2\pi p a) K_{s-\frac{1}{2}}(2\pi p
|c|) \right).
\eeq
Note that the expression (\ref{FirstSum}) is now valid even for
negative $s$. According to eqn.~(\ref{massivescalarperiodic}), we
need to differentiate eqn.~(\ref{FirstSum}) with respect to s and
set $s=-2$.  Since $\Gamma(-2) = \infty$, the derivative needs to
act only on $\Gamma(s)$ in eqn.~(\ref{FirstSum}). The sum over p,
when evaluated at $s=-2$, can be done exactly in terms of the
polylogarithm functions
\beqa
{\rm Li}_{n}(x) = \sum_{k=1}^{\infty} \frac{x^{k}}{k^{n}}\;.
\nonumber
\eeqa
The result is:
\beqa
\label{derivative1}
\left. \frac{d}{ds} F(s;a,c) \; \right|_{s=-2} =
-\frac{16\pi}{15} |c|^5 + \frac{1}{2\pi^4}
\left(4 \pi^2 |c|^2 {\rm Li}_3(q) + 6 \pi |c| {\rm Li}_4(q) + 3 {\rm Li}_5(q) + h.c.\right)~,
\eeqa
where we used $\Gamma\left(\frac{5}{2}\right) =
-\frac{8\sqrt{\pi}}{15}$ and  defined $q = e^{2 \pi i (a + i|c|)}$.
Setting $a=0$ or $a = \frac{1}{2}$ (for periodic and antiperiodic
boundary conditions respectively) and replacing in
eqn.~(\ref{massivescalarperiodic}) gives
eqn.~(\ref{massivepotential}) of Section \ref{5dcasimirmassive}
(the first term in eqn.~(\ref{derivative1}) corresponds to a finite
contribution to the bulk cosmological constant and was omitted in
eqn.~(\ref{5dcasimirmassive}), since other infinite contributions
have already been discarded by the regularization procedure).

Next, we turn to the double sum needed in the 6-d models, discussed
in Section \ref{6dcasimir}:
\beq
\sum_{m,n}{'} \frac{1}{|n + m \tau|^{2s}}~,
\eeq
where the prime indicates that the zero mode is to be excluded and
$\tau = \tau_{1}+ i \tau_{2}$. Writing the sum as:
\beqa
\sum_{m}{'}\sum_{n}\frac{1}{|n + m \tau|^{2s}} +
\sum_{n}{'}\frac{1}{n^{2s}} =
\sum_{m}{'}\sum_{n}\frac{1}{[(n + m \tau_{1})^{2} + m^{2}
\tau_{2}^{2}]^{s}} + 2 \zeta(2s) \;,
\eeqa
we note that the sum over $n$ in the first term is of the form
eqn.~(\ref{SingleSum}), with $a = m \tau_{1}$ and $c = m \tau_{2}$.
Thus, using eqn.~(\ref{FirstSum}) we obtain:
\beqa
\sum_{m,n}{'} \frac{1}{|n + m \tau|^{2s}} &=& 2 \zeta(2s) +
\frac{\sqrt{\pi} \Gamma \left(s-\frac{1}{2}\right)}{\Gamma(s)}
|\tau_{2}|^{1-2s}\,2 \zeta(2s-1) + \nonumber \\
& & \mbox{} + \frac{8 \pi^{s}}{\Gamma(s)} |\tau_{2}|^{\frac{1}{2}-s}
\sum_{m=1}^{\infty}\sum_{p=1}^{\infty} \left(\frac{p}{m}\right)^{s-\frac{1}{2}}
\cos(2\pi p m \tau_{1}) K_{s-\frac{1}{2}}(2\pi p m |\tau_{2}|) \;.
\eeqa
Differentiating with respect to $s$, setting $s=-2$ and expressing
the sum over $p$ in terms of the polylogarithm functions as we did
in eqn.~(\ref{derivative1}), gives eqn.~(\ref{Vtau}) of Section
\ref{6dcasimir}.

\end{document}